\documentclass[prb,amsmath,superscriptaddress,citeautoscript,twocolumn,
showpacs,floatfix]{revtex4}
\usepackage{bm}
\usepackage{amssymb}
\usepackage{graphicx}
\usepackage{amsmath, amsthm, amssymb, graphicx}
\usepackage[usenames]{color}

\newcommand{\be}{\begin{equation}}
\newcommand{\ed}{\end{equation}}
\newcommand{\ber}{\begin{eqnarray}}
\newcommand{\edr}{\end{eqnarray}}
\newcommand{\bern}{\begin{eqnarray*}}
\newcommand{\edrn}{\end{eqnarray*}}
\newcommand{\bfl}{\begin{flalign}}
\newcommand{\efl}{\end{flalign}}
\newcommand{\tr}[1]{\underset{#1}{\mathrm{tr}}}

\newcommand{\ud}{\mathrm{d}}
\newcommand{\vc}[1]{\mathbf{#1}}

\newcommand{\set}[1]{\mathcal{#1}}

\begin{document}

\title{Partitioned Density Matrices and Entanglement Correlators }

\author{T. Cox}
\affiliation{Department of Physics and Astronomy, University
of British Columbia, 6224 Agricultural Rd., Vancouver, B.C., Canada
V6T 1Z1}
\affiliation{School of Mathematics and Statistics, Victoria University of Wellington,
P.O. Box 600, Wellington 6140, New Zealand}

\author{ P.C.E. Stamp}
\affiliation{Department of Physics and Astronomy, University
of British Columbia, 6224 Agricultural Rd., Vancouver, B.C., Canada
V6T 1Z1}
\affiliation{School of Mathematics and Statistics, Victoria University of Wellington,
P.O. Box 600, Wellington 6140, New Zealand}
 \affiliation{Pacific
Institute of Theoretical Physics, University of British Columbia,
6224 Agricultural Rd., Vancouver, B.C., Canada V6T 1Z1}

\date{{\small \today}}

\begin{abstract}

The density matrix of a non-relativistic quantum system, divided into $N$ sub-systems, is rewritten in terms of the set of all partitioned density matrices for the system. For the case where the different sub-systems are distinguishable, we derive a hierarchy of equations of motion linking the dynamics of all the partitioned density matrices, analogous to the ``Schwinger-Dyson" hierarchy in quantum field theory. The special case of a set of $N$ coupled spin-$1/2$ ``qubits" is worked out in detail. The equations are then rewritten in terms of a set of ``entanglement correlators", which comprise all the possible correlation functions for the system - this case is worked out for coupled spin systems. The equations of motion for these correlators can be written in terms of a first-order differential equation for an entanglement correlator supervector.

\end{abstract}


\maketitle


\section{ Introduction}
 \label{sec:intro}

\begin{figure}
\vspace{0.8cm}
\begin{center}
\includegraphics[width=8.5cm]{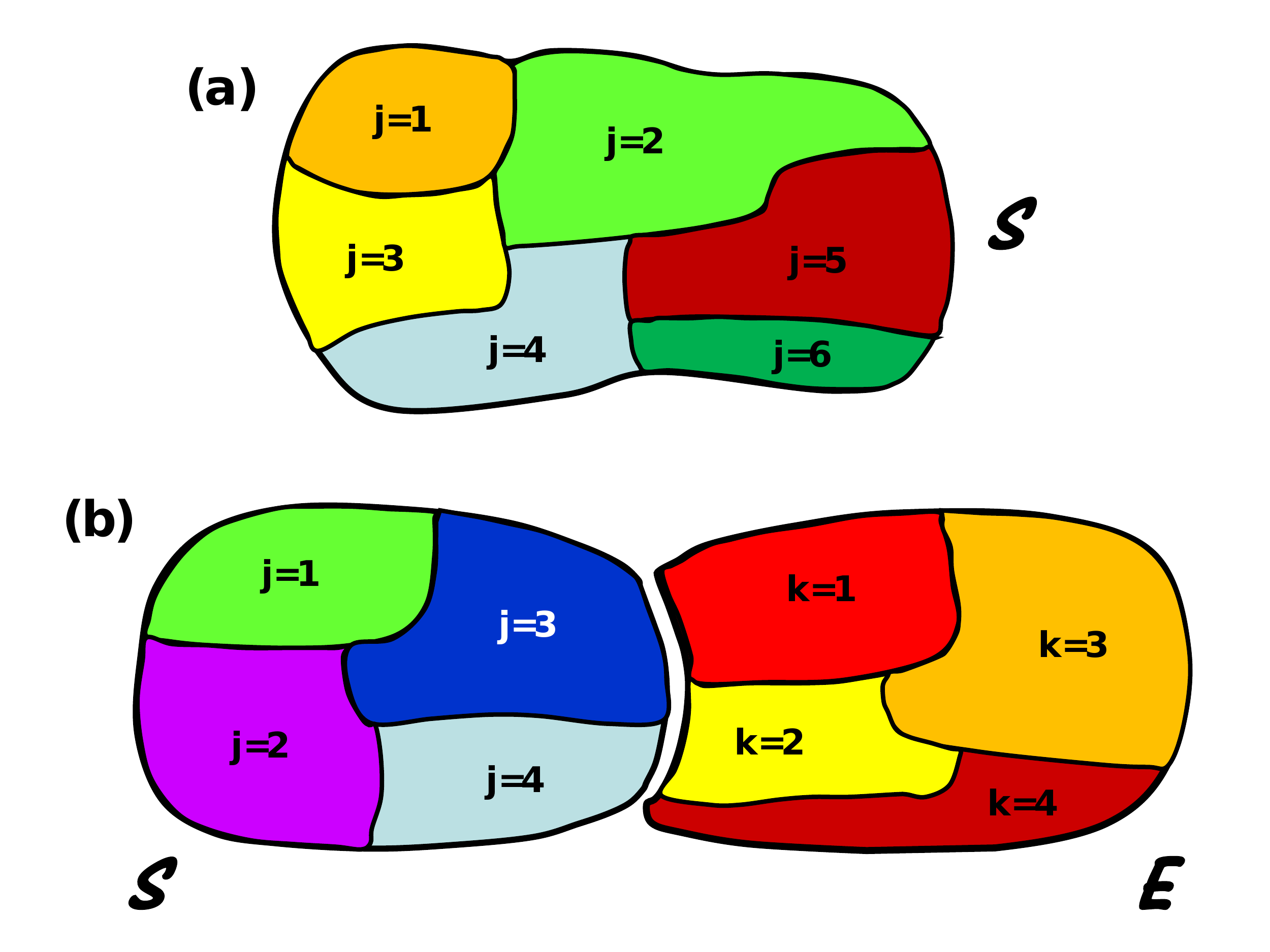}
\end{center}
\caption{ Partitioning of systems into cells; in (a) we show an isolated quantum system $\set{S}$ partitioned into $N=6$ subsystems, labelled by $j=1,\cdot, N$; in (b) we have a central system $\set{S}$ made up of $N = 4$ subsystems, labelled by $j=1,\cdot, N$, coupled to an environment $\set{E}$ consisting of $M = 4$ subsystems, labelled by $k=1,\cdot, M$.
 \label{fig:fig1}}
\end{figure}

In both physics and chemistry, the study of quantum-mechanical phenomena requires a definition of various statistical measures of correlation, between different sub-systems of a given physical system. Typically one is interested in two cases:

(i) We have an isolated system, and want to understand the internal correlations between its different parts, and their respective dynamics. Fig. \ref{fig:fig1}(a) shows a quantum system ${\cal S}$ with degrees of freedom divided into $N$ sub-systems $\sigma_j$, with $j=1,2,...N$. We wish to characterize the dynamics of ${\cal S}$ in terms of the dynamics of the correlations over the $N$ sub-systems \cite{statM}.

(ii) Our system ${\cal S}$ is coupled to an ``environment" ${\cal E}$, and we may wish to integrate out/average over at least some of the environmental degrees of freedom \cite{weiss99}. In Fig. \ref{fig:fig1}(b) we show ${\cal S}$ coupled to ${\cal E}$, which can itself be subdivided into $M$ sub-systems. Now we want to characterize the behaviour of both ${\cal S}$ and ${\cal E}$ in terms of both the internal correlations between their separate sub-systems, and also the correlations between sub-systems of ${\cal S}$ and ${\cal E}$. If we average wholly or partially over the environmental degrees of freedom, we would still like to be able to characterize the behaviour of ${\cal S}$.


In the first case, of an isolated system, this means we wish to find not only the dynamics of the total density matrix $\rho_{\cal S}(t)$ for ${\cal S}$, but also the dynamics of the reduced density matrices for all the different sub-sets of ${\cal S}$, and of the density matrices describing the correlations between these sub-sets. As an example, suppose that $N=3$. Then we can write $\rho^{\cal S}$ as
\begin{equation}
\rho_{\cal S} \;=\; \bar{\rho}_1 \bar{\rho}_2 \bar{\rho}_3 + \rho^C_{123} + \bar{\rho}_1 \bar{\rho}^C_{23} + \bar{\rho}_2 \bar{\rho}^C_{13} + \bar{\rho}_3 \bar{\rho}^C_{12}
 \label{rho123}
\end{equation}
where the $\bar{\rho}_j$ are reduced density matrices for the sub-system $j$, and the $\rho_{ij}^C$ and $\rho_{123}^C$ are density matrices describing the correlations between the sub-systems (including all the different kinds of multipartite entanglement between them). Our task is then to find the coupled equations of motion for each of these objects. In this paper we do this, both for the hierarchy of different correlated density matrices, and for a set of ``entanglement correlators" to be defined in the paper. We note that for $N$ sub-systems we will be dealing with all the different possible partitions of these subsets \cite{partition}.

One can treat both relativistic fields and non-relativistic systems of particles and/or spins, using the same general framework for each. However here we focus on non-relativistic systems, in which the sub-systems are distinguishable. As our primary example we will choose systems of spins or of qubits.

Two key questions that have driven this work are (i) what is a good way to characterize the many different levels of entanglement that exist between all the different sub-systems of a given $N$-body system; and (ii) what are the equations of motion for the coupled density matrices and the associated entanglement correlators, and how does this affect the dynamics of experimental quantities? The first question overlaps with work done over a long period in the quantum information community, notably on entanglement and separability  \cite{horodecki09,bennett96,wootters01,horedecki96,terhal00},
on measures of multipartite entanglement \cite{dur00,walter16,huber17}, on its detection \cite{guhne09,amico04}, and on ``disconnectivity" \cite{ajl80}.

However we stress that our main goal here has been to elucidate the second question. In this sense our results are - in a way to be explained - a generalization of the ``Schwinger-Dyson" hierarchy of coupled field correlators in either relativistic quantum field theory \cite{schwingerD} or non-relativistic $N$-particle theory \cite{martinS}. In the classical limit one can also see a relationship to the BBGKY hierarchy of equations of motion \cite{BBGKY}.

To fully appreciate the results to be described, one needs to develop the theory both for isolated systems like that in Fig. \ref{fig:fig1}(a), and for systems coupled to an environment like those depicted in Fig. \ref{fig:fig1}(b). To prevent this paper from becoming too long, we have divided the work into two parts: the present paper deals with isolated systems, and a following paper \cite{cox2} deals with systems coupled to a bath (where several new issues of principle arise). The main mathematical task of the present paper is the derivation of the relevant equations of motion. From a purely physical point of view, the present paper is is more relevant to the dynamics of entanglement, whereas the second paper focuses more on the dynamics of ``disentanglement" (ie., what is usually called ``decoherence").

Clearly there are many physical examples for which the results might be used. Here we have focused here on one specific system, a set of $N$ ``qubits", or spin-$1/2$ systems. In other papers we discuss  (i) specific applications to non-relativistic systems, notably the quantum Ising system, \cite{alvi18} and (ii) the treatment of relativistic quantum fields, where we reformulate the present work in terms of path integrals and ``composite field correlators" \cite{jordan18}.

We stress that in this paper we only derive the equations of motion, but do not try to solve them. To do this requires picking a specific Hamiltonian, and then - since any hierarchy of equations of motion is generally unsolvable - it needs some approximation scheme. In papers relying on this one, we have given approximate solutions for (a) the quantum Ising model, \cite{alvi18} and (b) the central spin model \cite{alvi19,Cspin}.

The plan of this paper is as follows. In section \ref{sec:rho+C}, we describe any non-relativistic many-body quantum system ${\cal S}$ in terms of a sum over all the different possible partitions of $n$ specific subsets of ${\cal S}$, of functions defined for each of these partitions. These different parts are assumed distinguishable. We then write this sum in terms of a complete set of ``entanglement correlated density matrices'' for the system. Then, in section \ref{sec:packet}, we show how this works for an $N$-qubit system, doing this for a pair and a triplet of qubits as well as for general $N$.

Moving on to dynamics, in section \ref{sec:EOM} we derive a hierarchy of coupled equations of motion for the partitioned density matrices over the different sub-systems, with the only assumption being pairwise interactions between these sub-systems; then we give results for the specific example of an $N$-qubit system. In section \ref{sec:PhysQ-B}, we connect all of this with the measurement of physical quantities, by defining an ``entanglement correlator supervector", which has as coordinates a list of {\it all possible} correlation functions that can be defined between {\it all} of the different operators which can act on the system. The equation of motion of this vector satisfies a first-order differential equation, and is simple to analyze.

Because the use of partitioned density matrices and their dynamics is rather novel, we have tried to keep technical details to a minimum in the main body of the paper (and stressed simple examples for the same reason). Accordingly, lengthy derivations have been relegated to several Appendices.


\section{Partitioned Density Matrices and their Correlations}
 \label{sec:rho+C}


In what follows we first define a set of correlated density matrices in terms of the full (unreduced) density matrix of the entire system ${\cal S}$ we are dealing with. To make intuitively clear what these correlated density matrices are, we discuss in some detail the example of a system partitioned into 4 sub-systems. Then we give a general expression for the correlated density matrices for some part ${\cal A}_n$ of the entire system containing $n$ sub-systems; and we discuss one of the key defining properties of the entanglement correlated density matrices.


\subsection{Definition of correlated density matrices}
 \label{sec:rho+C-1}

\begin{figure}
\begin{center}
\includegraphics[width=0.45\textwidth]{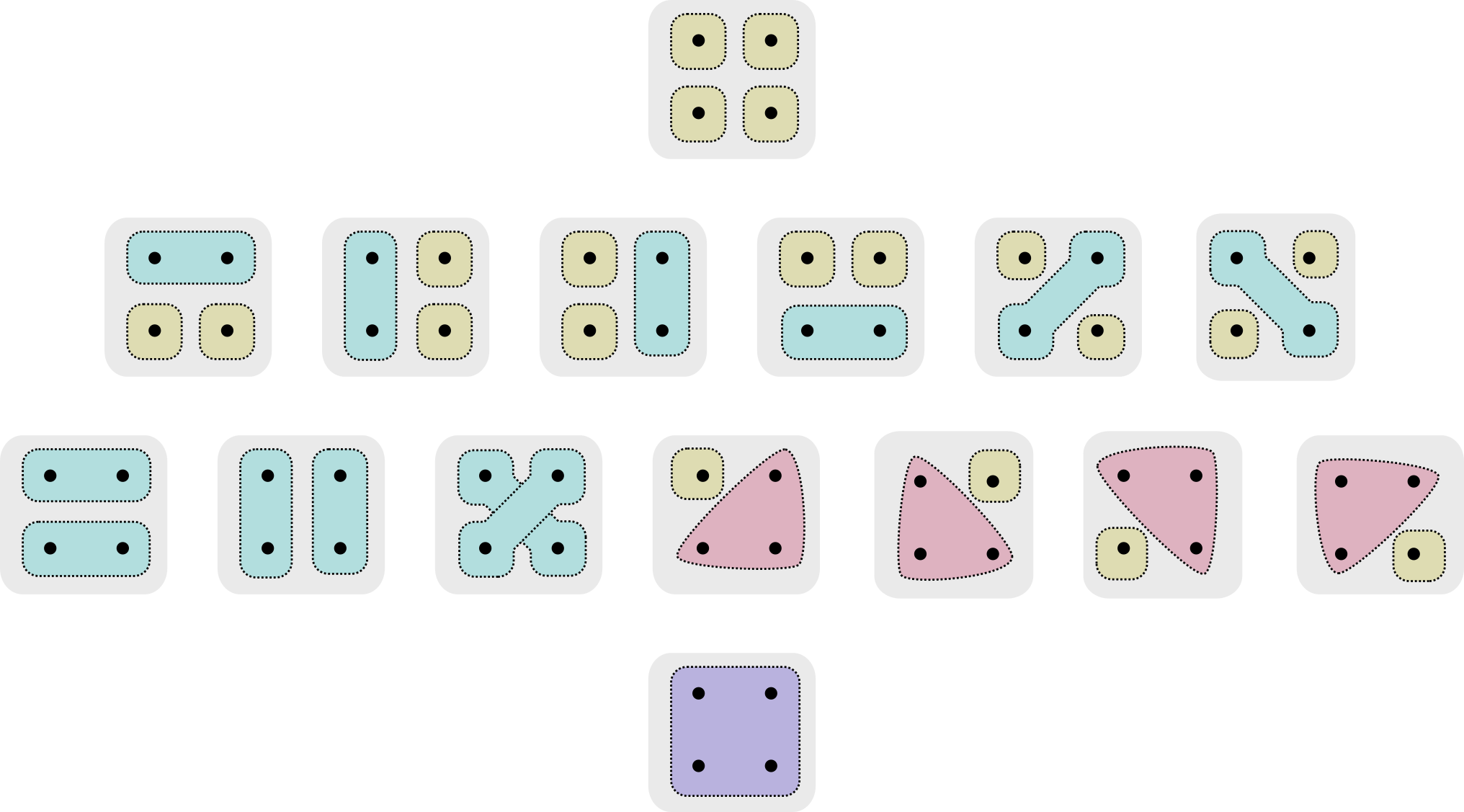}
\end{center}
\caption{The different partitions of a system containing $n=4$ distinguishable cells, with each cell denoted by a black dot. The total number of partitions is $B_4 = 15$, where $B_n$ is the Bell number. \label{fig:partitions4}}
\end{figure}

Consider a system $\mathcal{S}$ made up of some number $N$ of distinguishable disjoint subsystems (which we will often call ``elementary cells'', or just ``cells" for short). We may then enumerate all possible different ways of partitioning ${\cal S}$ into groups of subsets - this list constitutes a set $\mathfrak{P}_\set{S}$. As an example, in Fig. \ref{fig:partitions4} we show the various partitions for the case $N=4$. We can also enumerate all possible subsets of ${\cal S}$; this list forms another set ${\cal A}_S$. 

The two sets $\mathfrak{P}_\set{S}$ and ${\cal A}_S$ are not the same. Thus, suppose we have $N$ elementary cells. The set $\mathfrak{P}_\set{S}$ of all partitions of ${\cal S}$ then contains $B_N$ members, where $B_N$ is the Bell number \cite{bellN}; we will label the different members by $\mathfrak{p}_{\mu}$, with $\mu = 1,2,....B_N$, noting that one of the partitions $\mathfrak{p}_{\mu}$ contains only $\set{S}$ itself. The number $B_N$ grows super-exponentially with $N$ (we have $B_1 = 1, B_2 = 2, B_3 = 5, B_4 = 15, B_5 = 52$, and already $B_15 \sim 1.4 \times 10^9$). We will not, in this paper, need to know anything more about $B_N$.

The set ${\cal A}_S$, on the other hand, simply has as members the different subsets of $\mathcal{S}$; it is usually called the ``power set" of $\mathcal{S}$. If $\mathcal{S}$ has $N$ members, then the total number of members of ${\cal A}_S$ is just $2^N$; these are easily enumerated. We will label the members of ${\cal A}_S$ by $a_{\alpha}$, where $a_{\alpha} = 1,2,...2^N$, for a set $\mathcal{S}$ containing $N$ members.

Notice that any given partition of $\mathcal{S}$ is made up of a specific group of subsets of $\mathcal{S}$ (thus, eg., the partition $(12|3|4)$ of a set $\mathcal{S}$ of 4 members - depicted as the 2nd of the 15 members of the partitions of this set in Fig. \ref{fig:partitions4} - is made up of the subsets $(12)$, $(3)$, and $(4)$ of $\mathcal{S}$). We can write this statement as $\mathfrak{p}_{\mu} = \prod_{a_{\alpha} \in \mathfrak{p}_{\mu}} a_{\alpha}$. 

With these distinctions in mind, we would like in what follows to find an expression for the total density matrix of the system in terms of all the different reduced density matrices for the different subsets of $\mathcal{S}$, and of all the different entanglement correlated density matrices. 

We will give a precise definition of these entanglement correlated density matrices below. The reduced density matrices are defined in the usual way, ie., we define the reduced density matrix $\bar{\rho}_{a_{\alpha}}$ for some specific subset $a_{\alpha}$ of $\mathcal{S}$ as the partial trace of the full density matrix those other subsystem cells $i \not\in a_{\alpha}$. We shall write this definition as
\begin{equation}
\label{eq:redrho}
\bar{\rho}_{a_{\alpha}} \equiv \tr{\mathcal{S}\backslash \{ a_{\alpha} \} } \;\; \rho_\mathcal{S}
\end{equation}
where $\mathcal{S}\backslash \{ a_{\alpha} \}$ denotes the set containing all cells except those in the subset $a_{\alpha}$; here and from now on a bar over a density matrix indicates it is a reduced density matrix. 

We can then write the full density matrix in the form
\begin{equation}\label{eq:rhofull}
\displaystyle \rho_\mathcal{S}=\sum_{\mathcal{A}\subseteq \mathcal{S}}\left(\prod_{j\not\in\mathcal{A}}\bar{\rho}_{j}\right)\bar{\rho}^{C}_{\mathcal{A}}
\end{equation}
that is, as the sum over all subsets $\mathcal{A}$ of $\mathcal{S}$  (including the sets $\varnothing$ and $\mathcal{S}$) of a ``correlated part'' $\bar{\rho}^{C}_{\mathcal{A}}$  multiplied by the reduced density matrices $\bar{\rho}_{j}$ on those remaining individual cells not contained in ${\cal A}$. The above expression should be read with the following conventions:
\begin{align}
\bar{\rho}^C_\varnothing&=1\\
\prod_{j\in\varnothing}\bar{\rho}_j&=1\\
\bar{\rho}^C_i&=0
\label{sum-SS}
\end{align}
ie., we have that (i) the correlated part of the density matrix $\bar{\rho}_\varnothing^C$ over a set containing no cells is $1$; (ii) the product of the reduced density matrices taken over no cells is taken to be $1$; and (iii) the correlated part of the density matrix for a single cell is zero. Consider, for example, some system with a number $N > 3$ cells; and consider the terms in the sum \eqref{eq:rhofull}, in the cases where (i) $\set{A}=\varnothing$, (ii) $\set{A}=\{1,2,3\}$ and (iii) $\set{A}=\set{S}$. These terms are then given by
\begin{align}
\rho_{\mathcal{S}} \;\;&=\;\; \prod_{i\in\set{S}}\bar{\rho}_i \qquad\qquad\qquad\qquad (\set{A}=\varnothing)     \\
\rho_{\mathcal{S}} \;\;&=\;\;  \left( \prod_{i\not\in\{1,2,3\}}\bar{\rho}_i\right)\bar{\rho}^C_{123}  \qquad \; (\set{A}=\{1,2,3\})  \\
\rho_{\mathcal{S}} \;\;&=\;\;  \bar{\rho}^C_{\set{S}} \qquad\qquad\qquad\qquad \;\;\;\;(\set{A}=\set{S})
\end{align}
respectively.

There are 2 properties of the entanglement correlated parts $\bar\rho^C_\set{A}$ that make them useful. First, eqtn. \eqref{eq:rhofull} is a linear expansion of the full density matrix in terms of the $\bar\rho^C_\set{A}$. Second, we will take it as one of the defining conditions for the entanglement correlated density matrices that if we trace any single cell out of $\bar\rho^C_\set{A}$ we get zero; ie., we have for any $i\in\set{A}$ that
\begin{equation}\label{eq:trrhoc0}
 \tr{i} \; \bar{\rho}^{C}_\set{A}=0 \qquad\qquad (\forall i\in\set{A})
\end{equation}
Now eqtns. \eqref{eq:rhofull} and \eqref{eq:trrhoc0}, taken together, define the correlated parts $\bar\rho^C_\set{A}$ uniquely. However one needs to unpack these equations to see what they really mean; and we would also like to have an explicit expression for $\bar\rho^C_\set{A}$. In what follows we first see how to understand (\ref{eq:rhofull}) with simple examples; and we then find the desired expression for $\bar\rho^C_\set{A}$.


\subsection{A 4-cell example}
 \label{sec:rho+C-2}


The kind of thing we are talking about can be simply understood by looking at a system ${\cal S}$ composed of 4 sub-systems. In what follows we do this, introducing a diagrammatic representation of the results, and showing how the expansion over entanglement correlators can also be related to one over ``cumulant density matrices".

\subsubsection{Expansion over entanglement correlators}
 \label{sec:entCorr}

Let us begin by looking at only 2 sub-systems (what we will call a ``2-cell" system). The total density matrix $\rho^{\cal S}$ for ${\cal S}$ is then
\begin{equation}
\rho_{\cal S} \; \equiv \; \rho_{12} \;\;=\;\; \bar{\rho}_1 \bar{\rho}_2 + \rho^C_{12}
 \label{rho12}
\end{equation}
where $\bar{\rho}_1$ and $\bar{\rho}_2$ are the reduced density matrices for sub-systems $1$ and $2$ respectively, and $\rho^C_{12}$ is that part of $\rho_S$ in which there are correlations between the two sub-systems. We write $\rho_{\cal S} = \rho_{12}$ here to indicate the system now is just made up of two sub-systems $1$ and $2$. 

Notice that (\ref{rho12}) actually defines what we mean by $\rho^C_{12}$, ie., we have {\it defined} $\rho^C_{12}$ as
\begin{equation}
\rho_{12}^C=\bar{\rho}_{12}-\bar{\rho}_1\bar{\rho}_2.
\end{equation}
in terms of $\rho_{\cal S}$, $\bar{\rho}_1$, and $\bar{\rho}_2$. The generalization of (\ref{rho12}) to a 3-cell system is simple, and was given already above, in eq. (\ref{rho123}) of the introduction.

A system consisting of 4 sub-systems, whose partitions were already shown in Fig. \ref{fig:partitions4}, turns out to be more interesting. Then \eqref{eq:rhofull} reads
\begin{align}
\rho_{1234}\;=&\;\bar{\rho}_1\bar{\rho}_2\bar{\rho}_3\bar{\rho}_4
+\bar{\rho}_{12}^C\bar{\rho}_3\bar{\rho}_4+\bar{\rho}_{13}^C\bar{\rho}_2\bar{\rho}_4 + \bar{\rho}_{14}^C\bar{\rho}_2\bar{\rho}_3   \nonumber \\    &   +\bar{\rho}_{23}^C\bar{\rho}_1\bar{\rho}_4+
\bar{\rho}_{24}^C\bar{\rho}_1\bar{\rho}_3+\bar{\rho}_{34}^C\bar{\rho}_1\bar{\rho}_2 + \bar{\rho}_{123}^C\bar{\rho}_4  \nonumber \\    & +\bar{\rho}_{234}^C\bar{\rho}_1+\bar{\rho}_{134}^C\bar{\rho}_2+
\bar{\rho}_{124}^C\bar{\rho}_3+\rho^C_{1234}.
 \label{r1234}
\end{align}

Let us first notice how we get the lower reduced density matrices from this. We can immediately trace out cell $4$, to get $\bar\rho_{123}$; then, because $\tr{4}\bar{\rho}_{14}^C=\tr{4}\bar{\rho}_{24}^C=\ldots=\tr{4}\bar{\rho}^C_{124}=\ldots
=\tr{4}\bar\rho^C_{1234}=0$, we have
\begin{align}
 \label{eq:rho123}
\bar{\rho}_{123} \; \equiv &\; \tr{4}\rho_{1234} \nonumber \\ =&\; \bar{\rho}_1\bar{\rho}_2\bar{\rho}_3+
\bar{\rho}_3\bar{\rho}_{12}^C+\bar{\rho}_2\bar{\rho}_{13}^C+\bar{\rho}_1\bar{\rho}_{23}^C+
\bar\rho^C_{123}
\end{align}
which is just eqtn. (\ref{rho123}). We can then trace out cell $3$, as well, to get
\begin{equation}
\bar{\rho}_{12}\equiv\tr{\{3,4\}}\rho_{1234} \;\;=\;\; \bar{\rho}_1\bar{\rho}_2+\bar{\rho}_{12}^C.
\end{equation}
which is just eqtn. (\ref{rho12}).

Analogous expressions exist for $\bar{\rho}_{23}^C$ and $\bar{\rho}_{13}^C$; substituting these
into expression \eqref{eq:rho123} and rearranging we then find
\begin{equation}
\bar{\rho}_{123}^C=\bar{\rho}_{123}-\bar{\rho}_{12}\bar{\rho}_3-\bar{\rho}_{13}\bar{\rho}_2-
\bar{\rho}_{32}\bar{\rho}_1+2\bar{\rho}_1\bar{\rho}_2\bar{\rho}_3.
\end{equation}
so that finally we get an expression for the fourth order correlated part of the density matrix as
\begin{align}
\bar{\rho}_{1234}^C=&\rho_{1234}-\bar{\rho}_{123}^C\bar{\rho}_{4}-\bar{\rho}_{234}^C\bar{\rho}_{1}-
\bar{\rho}_{134}^C\bar{\rho}_{2}-\bar{\rho}_{1234}^C\bar{\rho}_{3}\nonumber\\&
+\bar{\rho}_{12}^C\bar{\rho}_3\bar{\rho}_4+\bar{\rho}_{13}^C\bar{\rho}_2\bar{\rho}_4+
\bar{\rho}_{14}^C\bar{\rho}_2\bar{\rho}_3+\bar{\rho}_{23}^C\bar{\rho}_1\bar{\rho}_4 \nonumber\\ &+
\bar{\rho}_{24}^C\bar{\rho}_1\bar{\rho}_3+\bar{\rho}_{34}^C\bar{\rho}_1\bar{\rho}_2-
3\bar\rho_{1}\bar\rho_{2}\bar\rho_{3}\bar\rho_{4}.
 \label{rho4-C}
\end{align}

At this point it is very useful to introduce a diagrammatic representation for the various functions involved. We represent the different cells or sub-systems with ``bullets'' (ie., by the symbol $\bullet$), and  the reduced density matrix for a group of cells is shown by linking these cells with a thick line. Then, for example, the expression $\bar\rho_{134}\bar\rho_2$ is represented as shown in Fig. \ref{fig:rho3}(a). 

We now represent the entanglement correlated density matrices, like $\bar\rho_{12}^C,\bar\rho_{123}^C,\ldots$, by double lines linking the cells. Then, in the 4-cell example, we have for the relation between the full density matrix $\rho_{1234}$ and the entanglement correlated density matrices $\rho^C$, given above in \eqref{r1234}, the diagrammatic representation shown in Fig. \ref{fig:rho4}.

Before continuing with the analysis, we remark two things about these results:

(i) we are not summing over different partitions to get these results, but over different subsets of the 4-site system, ie., over the power set. 

(ii) the number of different terms shown in Fig. \ref{fig:rho4} is not $2^4 = 16$, as one might naively expect for the power set of our 4-site system. Instead it is $2^4 - 4 = 12$. This is because the 4 subsets made from single individual sites gives no contribution - the correlated part of a single site density matrix is zero, as specified in eqtn. (\ref{sum-SS}). Thus in general we expect a total number of diagrams $2^N-N$ to contribute to the expansion (\ref{eq:rhofull}).

\subsubsection{Expansion over Cumulant matrices}

As just noted, the expansion (\ref{eq:rhofull}) is not an expansion over the different partitions of the total set $\mathcal{S}$, but over the power set. However one can also do an expansion defined directly in terms of these partitions, rather than by the zero trace condition in eqtn. \eqref{eq:trrhoc0}.

Suppose we take the set ${\mathfrak P}_\set{S}$ of all partitions of ${\cal S}$, and then for each one of these partitions we factorize the result into reduced density matrices for single cells uncorrelated with the rest, and a set of ``cumulant reduced density matrices'' $\rho^{CC}$ for the other cells. The expansion of the total density matrix in terms of these cumulant matrices then has the same structure as a cumulant expansion of a joint probability function or functional; ie., we can write
\begin{equation}\label{eq:rhorhocumu}
\rho_\set{S}=\sum_{\mathfrak{p}_\mu\in\mathfrak{P}_\set{S}}\prod_{\set{A}\in \mathfrak{p}_\mu}\bar{\rho}^{CC}_\set{A}.
\end{equation}
Equation \eqref{eq:rhorhocumu} can be used to inductively to define $\bar{\rho}^{CC}_\set{A}$, with the convention that for a single elementary subsystem, the cumulant matrix $\bar{\rho}_i^{CC}$ is defined to be the reduced density matrix, ie., $\bar{\rho}_i^{CC}\equiv\bar{\rho}_i$.


\begin{figure}
\begin{center}
\includegraphics[scale=1]{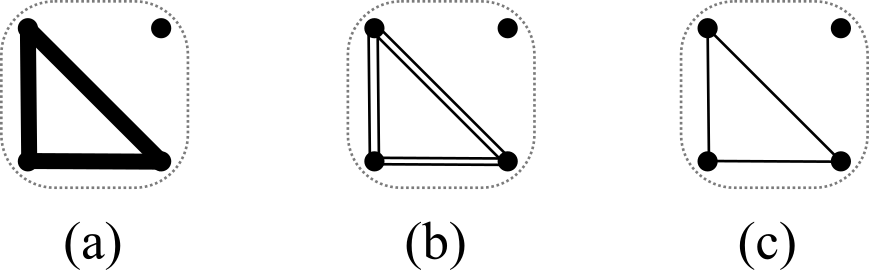}
\end{center}
\caption{Diagrammatic representation of some of the terms in the 4-cell density matrix. In (a) we show the term  $\bar\rho_{134}\bar\rho_2$ appearing in eqtn. (\ref{rho4-C}); in (b) we show the term $\bar\rho_{134}^C\bar\rho_2$, also appearing in eqtn. (\ref{rho4-C}); and in (c) we show the term  $\bar\rho_{134}^{CC}\bar\rho_2$, appearing in eqtn. (\ref{eq:rhointermsofrhocc}).
\label{fig:rho3}}
\end{figure}


The relation between this cumulant expansion and the power set expansion we are using here, which is given in terms of entanglement correlated density matrices, is easily illustrated for the 4-cell problem, for which we find the cumulant expansion 
\begin{align}\label{eq:rhointermsofrhocc}
\rho_{1234}=&\bar{\rho}_{1234}^{CC}+\bar{\rho}_{123}^{CC}\bar{\rho}_4+
\bar{\rho}_{124}^{CC}\bar{\rho}_3+\bar{\rho}_{134}^{CC}\bar{\rho}_2+
\bar{\rho}_{234}^{CC}\bar{\rho}_1
 \nonumber\\
&+\bar{\rho}_{12}^{CC}\bar{\rho}^{CC}_{34}+\bar{\rho}_{14}^{CC}
\bar{\rho}^{CC}_{23}+\bar{\rho}_{13}^{CC}\bar{\rho}^{CC}_{24}+
\bar{\rho}_{23}^{CC}\bar{\rho}_1\bar{\rho}_4
 \nonumber\\
&+\bar{\rho}_{13}^{CC}\bar{\rho}_2\bar{\rho}_4+\bar{\rho}_{14}^{CC}
\bar{\rho}_2\bar{\rho}_3+\bar{\rho}_{12}^{CC}\bar{\rho}_3\bar{\rho}_4+
\bar{\rho}_{24}^{CC}\bar{\rho}_1\bar{\rho}_3
 \nonumber\\&+
\bar{\rho}_{34}^{CC}\bar{\rho}_1\bar{\rho}_2+
\bar{\rho}_1\bar{\rho}_2\bar{\rho}_3\bar{\rho}_4
\end{align}
for $\rho_{1234}$ in terms of the $\rho^{CC}$.

One can of course invert the relation (\ref{eq:rhorhocumu})  as well. Thus, for example, the 4th-order cumulant density matrix is given in terms of the entanglement correlated matrices $\rho^C$ and the reduced density matrices by
\begin{equation}
\rho^{CC}_{1234}\;\equiv\; \bar{\rho}_{1234}^C-\rho^C_{12}\rho^C_{34}
-\rho^C_{14}\rho^C_{23}-\rho^C_{13}\rho^C_{24}
\end{equation}
which when expanded out gives
\begin{align}
\rho^{CC}_{1234}\;\;\equiv&\;\; \bar{\rho}_{1234}-\bar{\rho}_{123}^C\rho_{4}-\bar{\rho}_{234}^C\rho_{1}-
\bar{\rho}_{134}^C\rho_{2}-\bar{\rho}_{1234}^C\rho_{3} \nonumber\\ &  -\bar\rho^C_{12}\bar\rho^C_{34} -
\bar\rho^C_{14}\bar\rho^C_{24}-\bar\rho^C_{13}\bar\rho^C_{24} \nonumber\\&
+2(\bar{\rho}_{12}^C\bar{\rho}_3\bar{\rho}_4+\bar{\rho}_{13}^C\bar{\rho}_2\bar{\rho}_4+
\bar{\rho}_{14}^C\bar{\rho}_2\bar{\rho}_3+\bar{\rho}_{23}^C\bar{\rho}_1\bar{\rho}_4+  \nonumber\\ &
\bar{\rho}_{24}^C\bar{\rho}_1\bar{\rho}_3+\bar{\rho}_{34}^C\bar{\rho}_1\bar{\rho}_2) \;-\;
6\bar\rho_1\bar\rho_2\bar\rho_3\bar\rho_4.
\end{align}

We can also illustrate the cumulant expansion diagramatically.
If we represent the cumulant reduced density matrices $\bar\rho_{12}^{CC},\ldots$ by single lines between the relevant cells (compare Figs.  \ref{fig:rho3}(b) and  \ref{fig:rho3}(c)).
Then, for the relation between the full density matrix $\rho_{1234}$ and the cumulant density matrices $\rho^{CC}$, we have the diagrammatic representation shown in Fig. \ref{fig:rho4C}.


\begin{figure}
\begin{center}
\includegraphics[width=8.5cm]{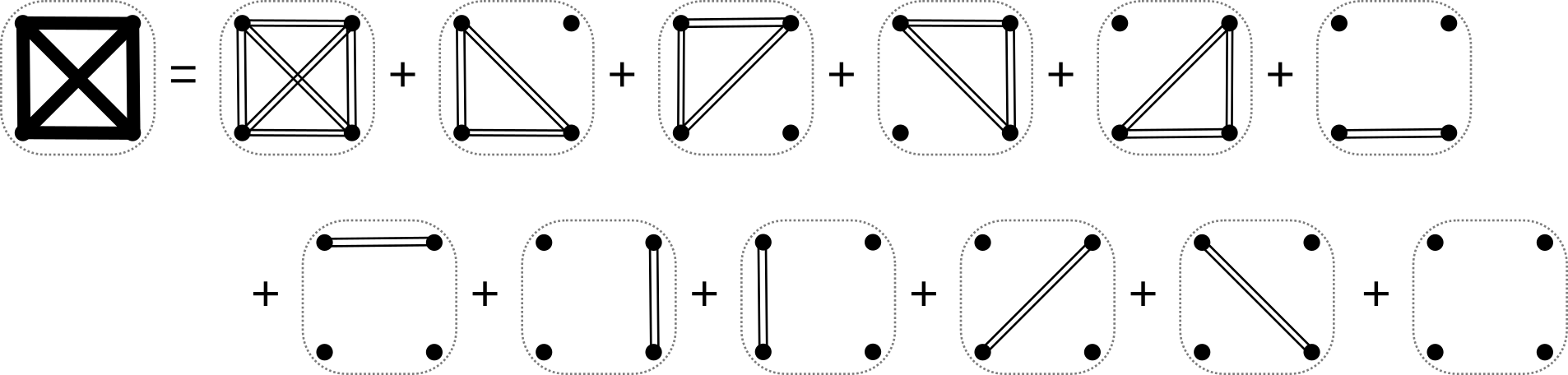}
\end{center}
\caption{Diagrammatic representation of the expansion of $\rho_{1234}$ into density matrices for the four sub-systems, as expressed in eqtn. \eqref{r1234}}
 \label{fig:rho4}
\end{figure}


We see that the relationship between the full density matrix $\rho_{\cal S}$ and the cumulant density matrices $\rho^{CC}$ is the same as that in a typical cumulant expansion, and so can be derived in the usual way for any value of $n$.


\begin{figure}
\begin{center}
\includegraphics[width=8.5cm]{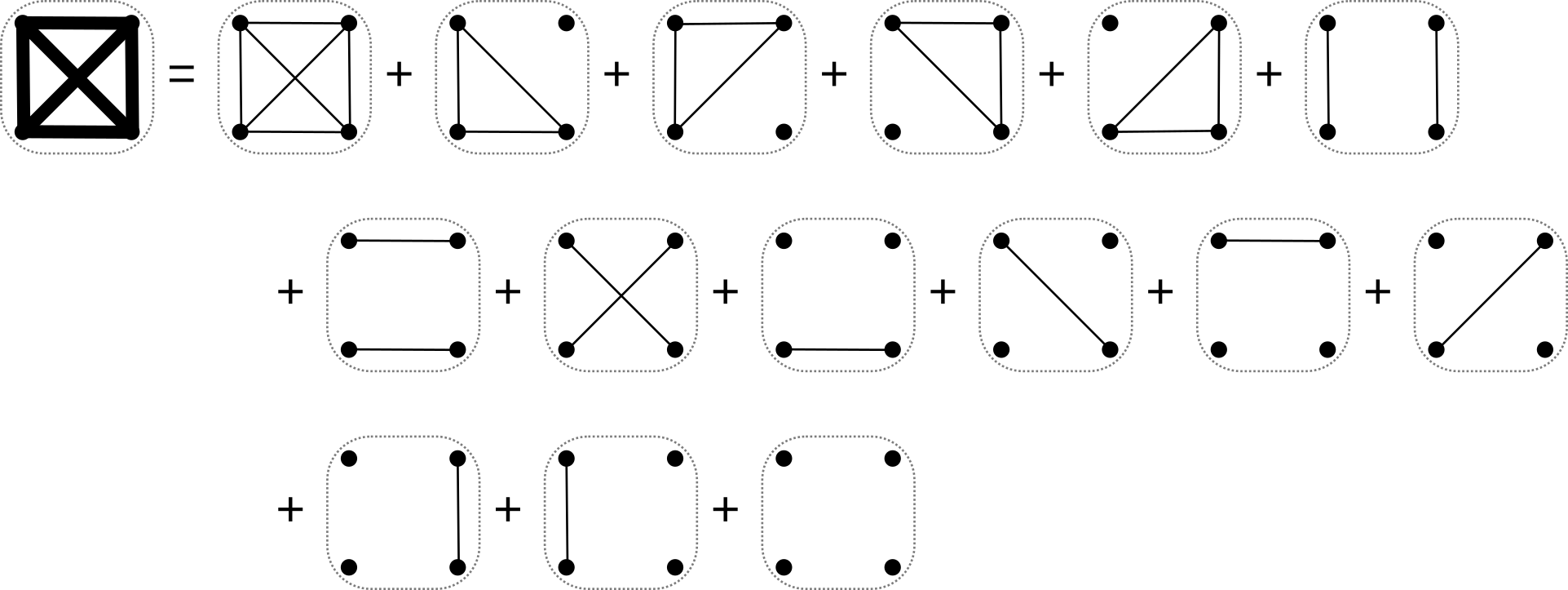}
\end{center}
\caption{Diagrammatic representation of the expansion of the $4$-cell density matrix into the $B_4 = 15$ different cumulant density matrices for the sub-systems, given in eqtn. \eqref{eq:rhointermsofrhocc}}
 \label{fig:rho4C}
\end{figure}


\subsection{General properties of entanglement correlated density matrices}
 \label{sec:rho+C-3}

As we have just seen, the relationship between $\rho_{\cal S}$ and the cumulant density matrices $\rho^{CC}$ is relatively straightforward. On the other hand, the relationship between $\rho_{\cal S}$ and the entanglement correlated density matrices $\rho^C$ is not so obvious - we still do not have a general expression for the correlated part of the total density matrix. To properly understand things we now turn to the general case.

\begin{figure}
\vspace{-0.7cm}
\begin{center}
\includegraphics[width=8.5cm]{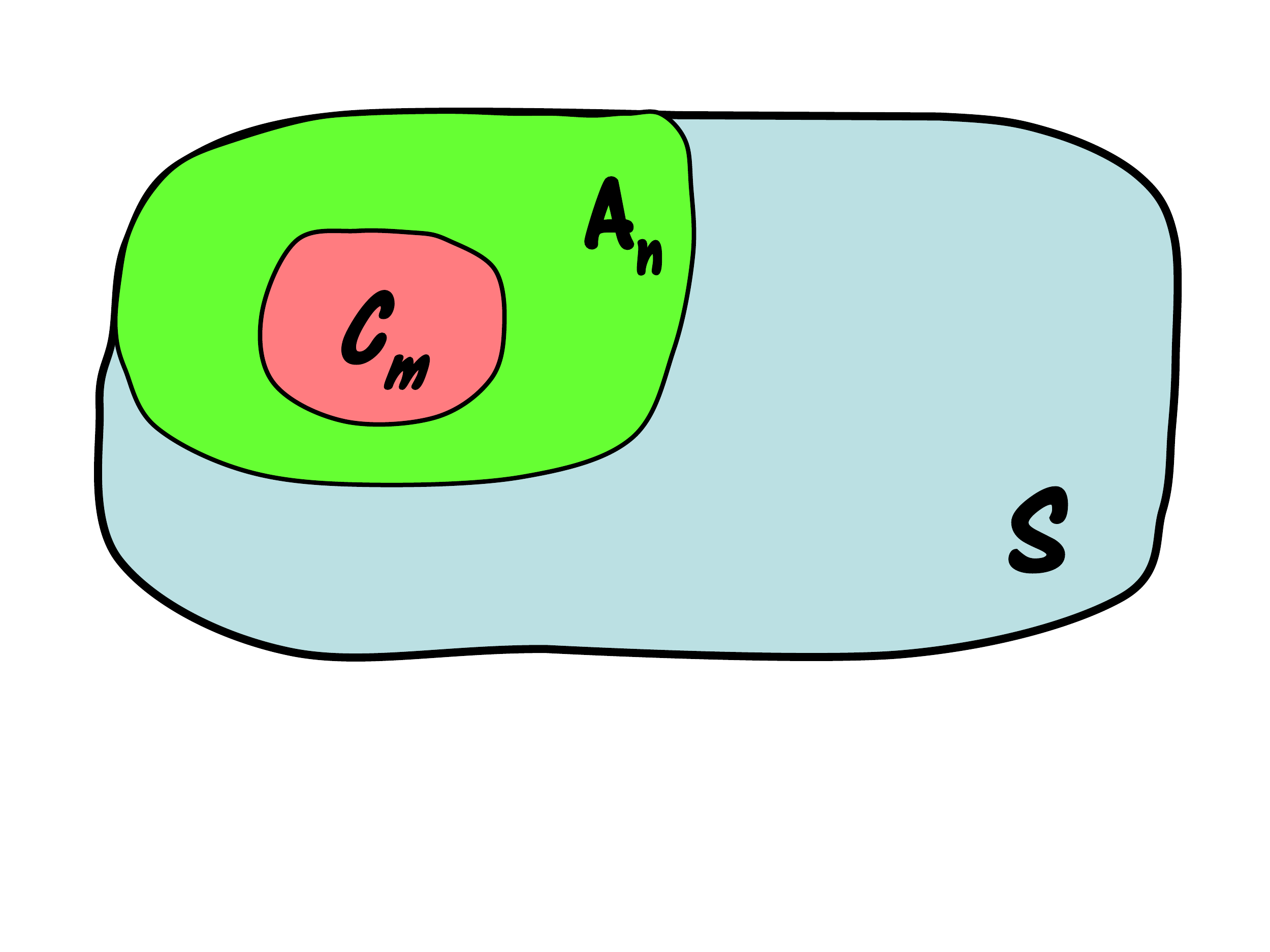}
\end{center}
\vspace{-2cm}
\caption{A representation of the sets used in equation \eqref{eq:rhoC}. The set $\set{A}_{n}$ is a subset of the whole system $\set{S}$, and contains $n$ members. The set $\set{C}_{m}$, which contains $m$ members, is a subset of $\set{A}^{n}$.}
 \label{fig:SSA6}
\end{figure}

What we wish to show is how, for a general subset ${\mathcal{A}_{\alpha}^{(n)}}$ of $n$ cells of a total system ${\cal S}$ containing $N$ cells, the correlated part of the reduced density matrix can be written as a sum over terms involving the reduced density matrices for all subsets $\mathcal{C}_{\mu}^{(m)}\subseteq\mathcal{A}_{\alpha}^{(n)}$. The notation used here labels the specific subsets $\mathcal{C}_{\mu}^{(m)}$ and $\mathcal{A}_{\alpha}^{(n)}$ by the subscripts $\mu$ and $\alpha$; the superscripts $m$ and $n$ tell us how many cells are contained in these subsets. This is illustrated in Fig. \ref{fig:SSA6}. The key result we find can be written as
\begin{align}
 \label{eq:rhoC-1}
 \bar\rho^{C}_{\mathcal{A}_{\alpha}^{(n)}}\;=&\;\sum_{m=2}^n(-1)^{(n-m)}
 \sum_{\mathcal{C}_{\mu}^{(m)}\subseteq\mathcal{A}^{(n)}_\alpha}\left(\bar\rho_{\mathcal{C}_m}\prod_{j\in \mathcal{A}_{\alpha}^{(n)}\backslash \mathcal{C}_{\mu}^{(m)}}\bar\rho_j\right) \nonumber\\ &  \qquad -(-1)^n(n-1)\prod_{j\in  \mathcal{A}_{\alpha}^{(n)}}\bar\rho_j.
\end{align}
which says that the entanglement correlated density matrix $\bar\rho^{C}_{\mathcal{A}_{\alpha}^{(n)}}$ for the specific set $\mathcal{A}_{\alpha}^{(n)}$ of cells can be written as a sum over entanglement correlated density matrices for all the different subsets $\mathcal{C}_{\mu}^{(m)}$ of $\mathcal{A}_{\alpha}^{(n)}$, multiplied by the product of the reduced matrices for all the cells $j$ that are not included in the subset $\mathcal{C}_{\mu}^{(m)}$ (this being the first term in \eqref{eq:rhoC-1}), minus a term which is simply the product of all the individual cell reduced density matrices for all the cells in $\mathcal{A}_{\alpha}^{(n)}$.

To reduce somewhat the profusion of indices in this expression, we will henceforth write expressions of this kind without the Greek indices labelling the specific subsets - thus \eqref{eq:rhoC-1} becomes
\begin{align}
\label{eq:rhoC}
 \bar\rho^{C}_{\mathcal{A}_n} \;=&\; \sum_{m=2}^n(-1)^{(n-m)}\sum_{\mathcal{C}_m\subseteq\mathcal{A}_n}
 \left(\bar\rho_{\mathcal{C}_m}\prod_{j\in \mathcal{A}_n\backslash \mathcal{C}_m}\bar\rho_j\right)  \nonumber\\ & \qquad -(-1)^n(n-1)\prod_{j\in  \mathcal{A}_n}\bar\rho_j.
\end{align}


The simplest way to demonstrate the result in eqtns. \eqref{eq:rhoC-1} and/or \eqref{eq:rhoC} is to construct an inductive proof - this is done in Appendix \ref{sec:App-A}. This result shows how one can define $n$-cell entanglement explicitly in terms of all possible combinations of $m$-cell entanglements over the different subsets of the $n$ cells, $\forall m < n$, along with products of single cell reduced density matrices. We shall see in the next two sections how we can employ eq.\eqref{eq:rhoC} to define a set of correlation functions which exhaustively characterize all the different kinds of entanglement that exist at the $n$th level, i., for a set of $n$ entangled cells.

As noted above, a key property of the entanglement correlation density matrices $\bar{\rho}^C$ is that any partial trace over $\bar{\rho}^C_{\set{A}_n}$ in \eqref{eq:rhoC}, ie., one in which we trace out any $i\in\set{A}_n$, gives zero - compare eqtn. (\ref{eq:trrhoc0}). In the discussion above, we treated this equation as a defining property of the $\bar{\rho}^C$. However, one can also derive the result explicitly from the expression \eqref{eq:rhoC}. The derivation is given in Appendix \ref{sec:App-A2}.

Let us now recapitulate. The basic result of this section is contained in eqtn. \eqref{eq:rhoC-1}, otherwise written as  \eqref{eq:rhoC}, which gives a way of decomposing a density matrix for some system ${\cal S}$ into a sum over reduced density matrices and correlated density matrices over all the possible sub-systems of ${\cal S}$. The discussion has been rather abstract. In the next section we see how to understand this result for some simple examples; and in section \ref{sec:EOM} we see how it may be applied to unravel the coupled dynamics of all the different sub-systems.


\section{Example: N-Qubit Spin System}
\label{sec:packet}


The example of a set of spins is extremely useful in understanding what is going on in the hierarchy of density matrices. In what follows we consider a system ${\cal S}$ of $N$ spin-$1/2$ ``qubits", with fixed pairwise interactions between them. This is a simple toy model for a quantum information processing system. It is also used to model many real physical materials devised for quantum information processing, where a decohering bath also exists - examples abound in solid-state electronic and nuclear spin systems \cite{N+V,semiC,SMM,LiHo} as well as neutral cold atoms \cite{coldA}.

In this example, our ``elementary cells" become much simpler - each cell is a single spin-$1/2$ degree of freedom. Because these   cells are irreducible, ie., can no longer be split into a set of smaller ``sub-cells", we will refer in this case to the cells as ``sites".

Apart from discussing the general $N$-qubit case, we also look in detail at pairs and triplets of spins ($N =2,3$). The results are useful - in particular, they teach us that the easiest way to understand the hierarchy of entanglement at the level of different qubits is just to look at the different partitioned correlated density matrices.

\subsection{General Results for $N$ coupled Qubits}
 \label{sec:Nqubit}

In what follows we wish to write some of the results of the last section for a set of $N$ qubits - these results will hold regardless of what kinds of interaction may exist between the qubits, or what external fields may be acting on them.

\subsubsection{Spin Representations}
 \label{sec:spinRep}

We begin by establishing some notation. In dealing with a set of  $N$ spin half's we write Pauli matrices for each spin as $\{\sigma_i^\mu\}$ (where $i\in\{1,2,\ldots,N\}$ labels the site and in the ``Cartesian" representation $\mu\in\{x,y,z\}$ denotes the Cartesian components). We will also use a ``ladder operator" representation: defining
\begin{equation}
\sigma^\pm\equiv\tfrac{1}{\sqrt{2}}\left(\sigma^x\pm i\sigma^y\right).
\end{equation}
we will use barred indices, $\bar{\mu}\in\{+,-,z\}$ to describe the different components of the spins in this representation so that
\begin{equation}
\langle\sigma^{\bar{\mu}}\rangle=\begin{pmatrix}
\langle \sigma^{+}\rangle\\
\langle\sigma^{-}\rangle\\
\langle\sigma^z\rangle
\end{pmatrix}.
\end{equation}

Then for a single spin we have the density matrix in the Bloch representation \cite{fano57}
\begin{equation}
\rho=\frac{1}{2}\left(1+\langle \boldsymbol{\sigma} \rangle\cdot\boldsymbol{\sigma}\right).
\end{equation}
so that $\tr{}\rho^2=\frac{1}{2}(1+\langle\boldsymbol{\sigma}\rangle^2)$, and for a pure state the polarization $\langle\boldsymbol{\sigma}\rangle$ sits on the Bloch sphere, with $|\langle\boldsymbol{\sigma}\rangle| = 1$; otherwise $|\langle\boldsymbol{\sigma}\rangle| < 1$. In the ladder representation
this single-spin density matrix is written
\begin{equation}
\rho=\frac{1}{2}\left(1+\langle\sigma_i^+\rangle\sigma_i^-+
\langle\sigma_i^-\rangle\sigma_i^++\langle\sigma_i^z\rangle\sigma_i^z\right).
\end{equation}

Notice that in the Cartesian representation the trace  $\tr{}\sigma^\mu \sigma^\nu=2\delta^{\mu\nu}$, so that the coefficient of a given operator in any operator expansion is the expectation of that operator. This is no longer true in the ladder representation, ie., $\tr{}\sigma^{\bar{\mu}}\sigma^{\bar{\nu}}\neq2\delta^{\bar{\mu}\bar{\nu}}$. However if we introduce a notation in which we distinguish between upper and lower indices, and define the lowered vector as the Hermitian conjugate of the operator with raised index, ie.,
\begin{equation}
\sigma_{\bar{\mu}}\equiv\left(\sigma^{\bar{\mu}}\right)^\dagger
\end{equation}
so that $\sigma_\pm=\sigma^\mp$ and $\sigma_z=\sigma^z$, then we have the trace identity:
\begin{equation}
\tr{}\sigma^{\bar{\mu}}\sigma_{\bar{\nu}}=2\delta^{\bar{\mu}}{}_{\bar{\nu}}.
 \label{trID}
\end{equation}

In what follows we will denote the eigenstates of $\hat{\sigma}_z$ by $|\uparrow \rangle$, $|\downarrow \rangle$, so that
\begin{align}
|\uparrow\rangle\langle\uparrow|&=\tfrac{1}{2}(1+\sigma^z)\\
|\downarrow\rangle\langle\downarrow|&=\tfrac{1}{2}(1-\sigma^z)
 \label{updown}
\end{align}
and for a pure state at some angle $\phi$ in the $xy$-plane,
\begin{align}
\rho_{\sigma \sigma'} \;=&\;\tfrac{1}{2}(|\uparrow\rangle+e^{i\phi}|\downarrow\rangle)
(\langle\uparrow|+e^{-i\phi}\langle\downarrow|) \nonumber \\
&\; =\tfrac{1}{2}(1+\cos\phi\sigma^x+\sin\phi\sigma^y).
\end{align}
with $\sigma, \sigma' = \pm 1$ labelling the rows and columns of the density matrix.

\subsubsection{General Results for $N$ qubits}

We assume a system of $N$ qubits $\{ \boldsymbol{\sigma}_j \}$, with $j = 1,2,....N$. Let us write the density matrix for this system ${\cal S}$ in the form
\begin{equation}\label{eq:rhonhalfs}
\rho_{\mathcal{S}}=\frac{1}{2^N}\sum_{\mathcal{C}\subseteq\mathcal{S}}
\Bigl\langle\prod_{i\in\mathcal{C}}\sigma_i^{\mu_i}\Bigr\rangle
\prod_{i\in\mathcal{C}}\sigma_i^{\mu_i}.
\end{equation}
in which the density matrix contains contributions from all $2^N$ distinct subsets $\mathcal{C}$ of the set ${\cal S}$. The contribution to the density matrix from a given cluster $\mathcal{C}$ is determined by the correlation tensor for those spins contracted into a product of the Pauli matrices then multiplied by a normalisation factor.

Clearly $\rho_{\mathcal{S}}$, composed entirely of Pauli matrices, must be Hermitian. The trace of $\rho_{\mathcal{S}}$ comes from the contribution in which $\mathcal{C}$ is the empty set (because all the Pauli matrices are traceless) which is $2^{-N}\tr{}(I)=1$ as required. One can verify that $\tr{}\left(\sigma_1^\mu\rho_{\mathcal{S}}\right)=\langle\sigma_1^\mu\rangle$ etc. by using the relation $\sigma_1^\mu\sigma_1^\alpha=\delta^{\mu\alpha}I_1+i\epsilon^{\mu\alpha\gamma}\sigma_1^\gamma$ and using the traceless property of the Pauli matrices (so that any term in the sum which contains a Pauli matrix after it has been multiplied by $\sigma_1^\mu$ gives zero). In general the density matrix must be positive semidefinite, although this is a hard condition to get a handle on using the representation \eqref{eq:rhonhalfs}, as it depends on the spectrum of $\rho_\set{S}$. If $\rho_{\mathcal{S}}$  represents a pure state then $\rho_{\mathcal{S}}^2=\rho_{\mathcal{S}}$, which can be used to derive those relations among the correlation functions which hold for pure states (see section \ref{subs:2spinhalfs} below for examples). More generally we have
\begin{equation}
\tr{}\rho_{\mathcal{S}}^2=\frac{1}{2^N}
\sum_{\mathcal{C}\subseteq\mathcal{S}}
\Bigl\langle\prod_{i\in\mathcal{C}}
\sigma_i^{\mu_i}\Bigr\rangle
\Bigl\langle\prod_{i\in\mathcal{C}}\sigma_i^{\mu_i}\Bigr\rangle\leq 1.
\end{equation}

As noted above, there are $2^N$ possible $\mathcal{C}\subseteq\mathcal{S}$. When one takes the partial trace of \eqref{eq:rhonhalfs} we see that the expression for a reduced density matrix on a set $\mathcal{A}\subset\mathcal{S}$ containing $n$ spins is of the same form as (\ref{eq:rhonhalfs}), viz.,
\begin{equation}
\rho_{\mathcal{A}}=\frac{1}{2^n}\sum_{\mathcal{C}\subseteq\mathcal{A}}
\Bigl\langle\prod_{i\in\mathcal{C}}\sigma_i^{\mu_i}\Bigr\rangle
\prod_{i\in\mathcal{C}}\sigma_i^{\mu_i}.
\end{equation}

By using the trace identity (\ref{trID}) we also see that the representation of the density matrix in terms of barred variables is just
\begin{align}
\label{eq:rhonhalfscoh}
\rho_{\mathcal{S}} \;\;=&\;\; \frac{1}{2^N}\sum_{\mathcal{C}\subseteq\mathcal{S}}
\Bigl\langle
\prod_{i\in\mathcal{C}}\sigma_i^{\bar{\mu}_i}\Bigr\rangle
\prod_{i\in\mathcal{C}}\sigma_i{}_{\bar{\mu}_i}
 \nonumber \\
=&\;\;
\frac{1}{2^N}\sum_{\mathcal{C}\subseteq\mathcal{S}}
\Bigl\langle\prod_{i\in\mathcal{C}}\sigma_i{}_{\bar{\mu}_i}\Bigr\rangle
\prod_{i\in\mathcal{C}}\sigma_i^{\bar{\mu}_i}.
\end{align}
so that this representation of $\rho_\set{S}$ is identical in form to the Cartesian representation in (\ref{eq:rhonhalfs}).


\subsection{Some examples}
 \label{sec:spinEx}


The following simple examples are useful in that they not only illustrate much of the general theory discussed so far, but they also indicate some of the ways in which it can be further developed.

\subsubsection{A pair of spins}
\label{subs:2spinhalfs}

Consider a pair of spins $\boldsymbol{\sigma}_1,\boldsymbol{\sigma}_1$, for which the density matrix is \cite{fano83}
\begin{equation}\label{eq:rho12}
\rho_{12}=\frac{1}{4}\Bigl(1+ \sum_{j = 1,2}\langle\sigma_j{}_\mu\rangle\sigma_j{}^\mu+
\langle\sigma_1{}_\mu\sigma_2{}_\nu\rangle\sigma_1{}^\mu\sigma_2{}^\nu\Bigr)
\end{equation}
We can split this up to a correlated and uncorrelated part, according to
\begin{align}
\rho_{12} \;=&\;\rho_1\rho_2+\rho^C_{12} \nonumber \\
=&\; \frac{1}{4} \prod_j (1+ \langle\sigma_j{}_\mu\rangle\sigma_j{}^\mu) +
\frac{1}{4}\langle
\langle\sigma_1{}_\mu\sigma_2{}_\nu\rangle\rangle\sigma_1{}^\mu\sigma_2{}^\nu
\end{align}
where we have defined
\begin{equation}
\langle\langle\sigma_1{}_\mu\sigma_2{}_\nu\rangle\rangle=
\langle\sigma_1{}_\mu\sigma_2{}_\nu\rangle-
\langle\sigma_1{}_\mu\rangle\langle\sigma_2{}_\nu\rangle.
\end{equation}

Now $\rho_{12}$ is a $4\times 4$ hermitian matrix with unit trace, and as such has 16-1=15 free real parameters, viz., 3 components of $\langle\boldsymbol{\sigma}_1\rangle$ and $\langle\boldsymbol{\sigma}_2\rangle$ each, and 9 components of $\langle\sigma_1{}_\mu\sigma_2{}_\nu\rangle$. In the case of a single qubit in a pure state, the spin had to lie on the Bloch sphere. In the two-qubit case things are more complicated; for a pure state one requires $\rho_{12}^2=\rho_{12}$, which leads to the following constraints on the correlators,
\begin{align}
3&=\langle\boldsymbol{\sigma}_1\rangle^2+
\langle\boldsymbol{\sigma}_2\rangle^2+
\langle\sigma_1{}_\mu\sigma_2{}_\nu\rangle
\langle\sigma_1{}^\mu\sigma_2{}^\nu\rangle
 \label{eq:pure12I}\\
\langle\sigma_1^\mu\rangle&=
\langle\sigma_1^\mu\sigma_2^\beta\rangle\langle\sigma_2^\beta\rangle\\
\langle\sigma_2^\mu\rangle&=
\langle\sigma_1^\beta\sigma_2^\mu\rangle\langle\sigma_1^\beta\rangle\\
\langle\sigma_1^\mu\sigma_2^\nu\rangle&=
\langle\sigma_1^\mu\rangle\langle\sigma_2^\nu\rangle-
\tfrac{1}{2}\varepsilon^{\mu\alpha\lambda}
\varepsilon^{\nu\beta\gamma}
\langle\sigma_1^\alpha\sigma_2^\beta\rangle
\langle\sigma_1^\lambda\sigma_2^\gamma\rangle.
 \label{eq:pure12IV}
\end{align}

This gives $1+3+3+9=16$ constraint equations on the correlators for a pure state - obviously only 10 of these are independent, since there is a $6$-dimensional set of real numbers which describes the possible pure states (8 real numbers describe a 2-qubit ket $|\psi\rangle$, reduced by two by the requirements of normalization and the invariance of $\rho_{12}=|\psi\rangle\langle\psi|$ under phase rotations). For the pure state,
\begin{equation}
|\psi\rangle \;=\; \sum_{\sigma\sigma'} a_{\sigma\sigma'}e^{i\phi_{\sigma\sigma'}}|\sigma \sigma'\rangle
\end{equation}
where $\sigma, \sigma' = |\uparrow \rangle, |\downarrow \rangle$; the normalization condition is then $\sum_{\sigma\sigma'}a_{\sigma\sigma'}^2 \;=\; 1$.

For a general mixed state of two qubits, equations (\ref{eq:pure12I}-\ref{eq:pure12IV}) are replaced by a set of three independent inequalities, which ensure the positivity of the density matrix\cite{gamel16}. This reflects the fact that a mixed state density matrix requires $15$ independent real parameters (the 16 required to define an arbitary $4
\times 4$ hermitian matrix, minus one because the matrix must be traceless) rather than the eight required to define a pure state.

Of particular interest for qubit pairs are ``cat states'', which are fully entangled. An example of such a state is $|\Psi_2^C\rangle$ with wave-function and density matrix given by
\begin{align}
|\Psi_2^C\rangle&\equiv\tfrac{1}{\sqrt{2}}
\left(|\uparrow\uparrow\rangle+
e^{i\phi_{\downarrow\downarrow}}|\downarrow\downarrow\rangle\right)\\
|\Psi_2^C\rangle\langle\Psi_2^C|&=\frac{1}{4}
\Bigl(1+\cos\phi_{\downarrow\downarrow}
\left[\sigma_1^x\sigma_2^x-\sigma_1^y\sigma_2^y\right] \nonumber \\
& \qquad +\sin\phi_{\downarrow\downarrow}
\left[\sigma_1^y\sigma_2^x+\sigma_1^x\sigma_2^y\right]+\sigma_1^z\sigma_1^z\Bigr)
\end{align}
When we come to look at entanglement dynamics, it is then the correlated part of these functions which will interest us.

Let us now consider the relationship between $\rho_{12}^C$ and the different types of entanglement. There is some subtlety in this \cite{horodecki09}, especially in the case of mixed states. Consider, for instance, a mixed state which is an incoherent mixture of the state $|\uparrow\uparrow\rangle$, with spins are polarised in the $z$ direction, and the state  $|\rightarrow\rightarrow\rangle$, with both spins polarised in the $x$ direction, so that
\begin{align}
\rho_{12}=&\; \frac{1}{2}\left(|\uparrow\uparrow\rangle
\langle\uparrow\uparrow|+|\rightarrow\rightarrow\rangle
\langle\rightarrow\rightarrow|\right) \nonumber \\
=&\; \frac{1}{4}\left[1+\tfrac{1}{2}(\hat{x}+\hat{z})
\cdot(\boldsymbol{\sigma}_1+\boldsymbol{\sigma}_2)+
\tfrac{1}{2}\left(\sigma_1^x\sigma_2^x+\sigma_1^z\sigma_2^z\right)\right].
\end{align}
Now $\rho_{12}$ has non-zero correlation functions; we have
\begin{align}
\langle\langle\sigma_1^x\sigma_2^x\rangle\rangle \;&=\;\langle\langle\sigma_1^z\sigma_2^z\rangle\rangle \nonumber \\ & =\; -\langle\langle\sigma_1^x\sigma_2^z\rangle\rangle\;=\;
-\langle\langle\sigma_1^z\sigma_2^x\rangle\rangle=\frac{1}{4}.
\end{align}
On the other hand, since $\rho_{12}$ is an incoherent mixture of two separable states, it has zero entanglement of formation\cite{bennett96}. This is not the only measure of entanglement; and for a general mixed state the formulae for different entanglement measures may be quite complicated.

This example shows nicely that it makes sense to consider directly the set of correlators, instead of the different entanglement measures. Because the full set of 15 correlators completely specifies the density matrix, all information about entanglement between the pair of qubits is then contained in these correlators. Since any entanglement witness \cite{horedecki96,terhal00,guhne09} used to detect entanglement is necessarily a Hermitian operator, it follows that its expectation can also be written as a weighted sum over the correlators. Thus we can simply use the correlators themselves as the primary quantities, whose behaviour is to be determined.

\subsubsection{Three spins}

For a system with three spins, the general density matrix is written as a sum over correlators as
\begin{align}
\rho_{123}=&\frac{1}{8}\Bigl(1+ \sum_{j} \langle\sigma_1{}_\mu\rangle\sigma_j{}^\mu + \sum_{i<j}\langle\sigma_i{}_\mu\sigma_j{}_\nu\rangle\sigma_i{}^\mu\sigma_j{}^\nu \nonumber\\ &\qquad\qquad +\;\langle\sigma_1{}_\mu\sigma_2{}_\nu\sigma_3{}_\lambda\rangle
\; \sigma_1{}^\mu\sigma_2{}^\nu \sigma_3{}^\lambda\Bigr).
\label{eq:rho3spin}
\end{align}
We now have a number of different types of entangled state. Consider as an example the three different states
\begin{align}
|\Psi_3^a\rangle&=\tfrac{1}{\sqrt{2}}\left(|\uparrow\uparrow\downarrow\rangle+
|\downarrow\downarrow\downarrow\rangle\right)\\
|\Psi_3^b\rangle&=\tfrac{1}{\sqrt{2}}\left(|\uparrow\uparrow\uparrow\rangle+
|\downarrow\downarrow\downarrow\rangle\right)
\end{align}
and
\begin{align}
|\Psi_3^c\rangle&\equiv \tfrac{1}{\sqrt{3}}\sum_{\sigma_1, \sigma_2, \sigma_3} |\sigma_1 \sigma_2 \sigma_3 \rangle \; \delta [(\sum_{j}\sigma_j)+ 1] \nonumber \\
&= \tfrac{1}{\sqrt{3}} ( |\uparrow\downarrow\downarrow\rangle + |\downarrow\downarrow\uparrow\rangle +
|\downarrow\uparrow\downarrow\rangle )
\end{align}
For each of these states we can find the non-zero expectation values for the correlators in the density matrix representation \eqref{eq:rho3spin}. Consider first $|\Psi_3^a\rangle$, for which
\begin{align}
|\Psi_3^a\rangle:&\quad\langle\sigma_3^z\rangle=\langle\sigma_1^y\sigma_2^y\rangle=
\langle\sigma_1^x\sigma_2^x\sigma_1^z\rangle=\langle\sigma_1^z\sigma_2^z\sigma_1^z\rangle \;=\;-1
\nonumber\\
&\quad\langle\sigma_1^x\sigma_2^x\rangle=\langle\sigma_1^z\sigma_2^z\rangle=
\langle\sigma_1^y\sigma_2^y\sigma_2^z\rangle \;=\; 1
\end{align}
We that $|\Psi_3^a\rangle$ does not have 3-qubit entanglement, because we can write $|\Psi_3^a\rangle=\frac{1}{\sqrt{2}}\left(|\uparrow\uparrow\rangle+
|\downarrow\downarrow\rangle\right)\otimes|\uparrow\rangle$, and this is reflected in the fact that the correlated part of the three point function $\langle\langle\sigma_1\sigma_2\sigma_3\rangle\rangle=0$ is zero. However it does have 2-qubit entanglement and single qubit polarization.

Now consider the other two states, for which we have
\begin{align}
|\Psi_3^b\rangle:&\quad\langle\sigma_1^z\sigma_2^z\rangle=
\langle\sigma_1^z\sigma_2^z\rangle=
\langle\sigma_1^z\sigma_3^z\rangle=\langle\sigma_2^z\sigma_3^z\rangle
 \nonumber \\
& \qquad\qquad\qquad =\;
\langle\sigma_1^x\sigma_2^x\sigma_3^x\rangle=1    \nonumber\\
&\quad \langle\sigma_1^x\sigma_2^y\sigma_3^y\rangle=
\langle\sigma_1^y\sigma_2^x\sigma_3^y\rangle=\langle\sigma_1^y\sigma_2^y\sigma_3^x\rangle=-1
\end{align}
for the second state, and
\begin{align}
      |\Psi_3^c\rangle:&\quad\langle\sigma_i^z\rangle=\langle\sigma_i^z\sigma_j^z\rangle
=-\tfrac{1}{3},\,\langle\sigma_i^x\sigma_j^x\rangle=\langle\sigma_i^y\sigma_j^y\rangle=
\tfrac{2}{3}\nonumber\\
&\quad\langle\sigma_1^z\sigma_2^z\sigma_3^z\rangle=1,\,
\langle\sigma_i^x\sigma_j^x\sigma_\ell^z\rangle=
\langle\sigma_i^y\sigma_j^y\sigma_\ell^z\rangle=\tfrac{2}{3},\,
 \nonumber \\
&\quad\qquad (\text{for }i,j,\ell\text{ distinct }\in\{1,2,3\} ).
\end{align}
for the third state. Both $|\Psi_3^b\rangle$ and $|\Psi_3^c\rangle$ do have three qubit entanglement, as the correlated 3-qubit functions are non-zero (this especially obvious in the case of $|\Psi_3^b\rangle$,  which is the superposition of two terms, each of which is obtained from a triple spin flip of the other). Both states also have 2-qubit entanglement, and  $|\Psi_3^c\rangle$ also has single qubit polarisation. It can be shown that the states $|\Psi_3^b\rangle$ and $|\Psi_3^c\rangle$ are members of the only two different classes of fully entangled 3-qubit states \cite{dur00}, and all other fully entangled states can be obtained from them by local operations assisted with classical communication.

We observe that neither of the states $|\Psi_3^b\rangle$, $|\Psi_3^c\rangle$ has a full ``3-qubit entanglement'' in the way that $|\Psi_2^C\rangle$ has full ``2-qubit entanglement''. For $|\Psi_2^C\rangle$ all the single qubit correlators are zero, whereas for the 3-qubit system it is impossible for the following three conditions to hold at once:
\begin{align}
\langle\sigma_i^\mu\rangle&=0\quad\forall i\in\{1,2,3\}\label{eq:3qubitst1}\\
\langle\sigma_i^\mu\sigma_j^\nu\rangle&=0\quad\forall i\neq j\in\{1,2,3\}\\
\rho_{123}&\text{ represents a pure state.}\label{eq:3qubitst3}
\end{align}

To show this, we note that the first two conditions imply $\rho_{123}=\frac{1}{8}\left(I+\langle\sigma_1{}_\mu\sigma_2{}_\nu\sigma_3{}_\lambda
\rangle\sigma_1^\mu\sigma_2^\nu\sigma_3^\lambda\right)$. We can then calculate $\rho_{123}^2$ and we find that the ``$\sigma_1\sigma_2\sigma_3$'' component is
\begin{align}
\frac{1}{32}\langle\sigma_1{}_\mu\sigma_2{}_\nu\sigma_3{}_\lambda\rangle\sigma_1^\mu
\sigma_2^\nu\sigma_3^\lambda
 \;\neq \; \frac{1}{8}
\langle\sigma_1{}_\mu\sigma_2{}_\nu\sigma_3{}_\lambda\rangle
\sigma_1^\mu\sigma_2^\nu\sigma_3^\lambda
\end{align}
where the inequality $\neq$ holds for any non-zero value of $\langle\sigma_1{}_\mu\sigma_2{}_\nu\sigma_3{}_\lambda\rangle $. Thus the state can't be pure.

\subsubsection{N-qubit states}
\label{ssec:Nqubitrho}

There are still simple questions one can ask about $N$-qubit states; for example,
whether an analogue of the statements (\ref{eq:3qubitst1}-\ref{eq:3qubitst3}) be true when we have $N$ qubits. In other words, one can ask: does the $N$-qubit density matrix
\begin{equation}\label{eq:rhoN}
\rho_{12\ldots N}=\frac{1}{2^N}\left(I+\langle\sigma_1^{\mu_1}\sigma_2^{\mu_2}
\ldots\sigma_N^{\mu_N}\rangle\sigma_1^{\mu_1}\sigma_2^{\mu_2}\ldots\sigma_N^{\mu_N}\right)
\end{equation}
represent a valid pure state?  The answer is that this is true only if $N=1$ or $N=2$. For $N=3$ we have just seen that is not a pure state, and proofs for the non-existence of pure states of the form \eqref{eq:rhoN} for $N\geq 4$ are given in the literature \cite{huber17}, and refs therein.

When we deal with the full complexity of $N$-qubit states, it is hard to get very far in their analysis beyond simple statements of this kind. The number of possible partitions of the system becomes immense, growing super-exponentially as the Bell number, and to characterize the entanglement properties is clearly going to be very complicated. There is a large body of literature on the different types of multipartite entanglement, along with several reviews \cite{guhne09,horodecki09,walter16}.

However, again, even for $N$ spins, any observable witness we build to diagnose this entanglement can be expressed as a sum of different clusters of Pauli operators. Thus again it makes sense to go back to the study the dynamics of these correlators, in order to understand the dynamics of entanglement - this is perhaps the main lesson of the examples just examined. We therefore now turn to this dynamics.


\section{Dynamics of Partitioned Density Matrices}
 \label{sec:EOM}


One of our main objectives in this paper is to derive the dynamics of the entanglement correlated density matrices. For a system ${\cal S}$ made up from $N$ sub-systems or "cells", this means finding the equations of motion for each of the reduced density operators $\bar\rho_{\mathcal{A}_{\alpha}^{(n)}}$, as well as the correlated density operators $\bar\rho^C_{\mathcal{A}_{\alpha}^{(n)}}$,which describe the different sub-sets $\bar\rho_{\mathcal{A}\alpha}^{(n)}$ of ${\cal S}$. Now, unless the Hamiltonian for ${\cal S}$ is trivially non-interacting (ie., it consists of a simple sum of terms over each cell, with no interactions between the cells), it is clear that these equations of motion will actually couple the different $\bar\rho^{C}_{\mathcal{A}_{\alpha}^{(n)}}$, since any sub-set ${\mathcal{A}_n}$ will have interactions with cells not contained in that sub-set (unless of course $\bar\rho^{C}_{\mathcal{A}_n} = {\cal S}$). Thus we will end up with set of coupled equations of motion, which takes the form of a hierarchy of coupled differential equations.

In what follows we begin by deriving the hierarchy for a general closed system in which all interactions between the different cell subsystems are pairwise. Then, in order to see how things look for a specific example, we derive the hierarchy for the system of $N$ qubits discussed in the previous section, with a set of local fields on each qubit as well pairwise interactions between them.


\subsection{Result for $N$-partite system}
 \label{sec:EOM-site}


In the most common kind of Hamiltonian in physics, one has (i) a ``free'' or trivial part which only acts inside individual cells, along with (ii) an interacting part which contains pairwise terms between cells. The Hamiltonian then takes the form
\begin{equation}
 H_S \equiv H_S^0+H_S^I\;\;=\;\;\sum_{j\in\mathcal{S}}\left(H^0_i+\frac{1}{2}\sum_{i\neq j\in\mathcal{S}}V_{ij}\right)
 \label{H-S}
\end{equation}

We make no assumptions for the moment about the nature of the cells, or of the interactions between them, except those assumptions already noted in the Introduction, viz., that we refer to distinguishable sets of degrees of freedom for each cell (so that there are no "exchange terms" between cells), and the system is assumed non-relativistic.

The equation of motion for the system density matrix is
\begin{equation}
i\hbar\partial_t\rho_{\mathcal{S}}=[H_S,\rho_{\mathcal{S}}].
\end{equation}

Starting from this equation, and taking its trace over all cells except those contained in $\mathcal{A}_n$, one can derive an equation of motion for the reduced density matrix $\bar\rho_{\mathcal{A}_n}$ which takes the form
\begin{equation}
 \label{eq:rhoA}
i\hbar\partial_t\bar\rho_{\mathcal{A}_n}=\left[\bar{H}_{\mathcal{A}_n},
\bar\rho_{\mathcal{A}_n}\right]+
\sum_{\ell\not\in\mathcal{A}_n}\tr{\ell}\left(\sum_{j\in\mathcal{A}_n}
\left[V_{j\ell},\; \bar\rho_{\mathcal{A}_n\cup\{\ell\}}\right]\right).
\end{equation}
where we have defined an effective local Hamiltonian (ie., one entirely restricted to $\mathcal{A}_n$), by
\begin{equation}
\bar{H}_{\mathcal{A}_n} \;=\;\sum_{j\in\mathcal{A}_n}\left(H^0_i+\frac{1}{2}\sum_{i\neq j\in\mathcal{A}_n}V_{ij}\right)
 \label{HA-eff}
 \end{equation}

Although equation (\ref{eq:rhoA}) apparently has a fairly simple form, its derivation is actually quite lengthy, and we have found no way to shorten it. This derivation appears in appendix \ref{sec:App-B1}.

We can interpret (\ref{eq:rhoA}) by noting first that the time evolution of $\bar{\rho}_{\mathcal{A}_n}$ is determined both by the local Hamiltonian ${H}_{\mathcal{A}_n}$, acting solely on $\mathcal{A}_n$, and by the effect of interactions on all possible sets containing $\mathcal{A}_n$ along with one other member.

One can think of the local effective Hamiltonian as one in which all interaction terms act solely on pairs of cells within $\mathcal{A}_n$, ie., it is an ``internal" effective Hamiltonian for $\mathcal{A}_n$. The second "interaction" term in (\ref{eq:rhoA}) is then one in which $V_{jl}$ couples $\bar\rho_{\mathcal{A}_n}$ to "larger" reduced density matrices $\bar\rho_{\mathcal{A}_n\cup\{\ell\}}$ which involve not only all the cells in ${\mathcal{A}_n}$ but also one other cell $\ell$ from ${\cal S}$ that is outside ${\mathcal{A}_n}$; we then sum over all the different cells $\{ \ell \}$ that are outside ${\cal A}_n$. That there is only one other cell involved follows because we have only pairwise interactions in the original Hamiltonian.

\begin{figure}
\begin{center}
\includegraphics[width=8.5cm]{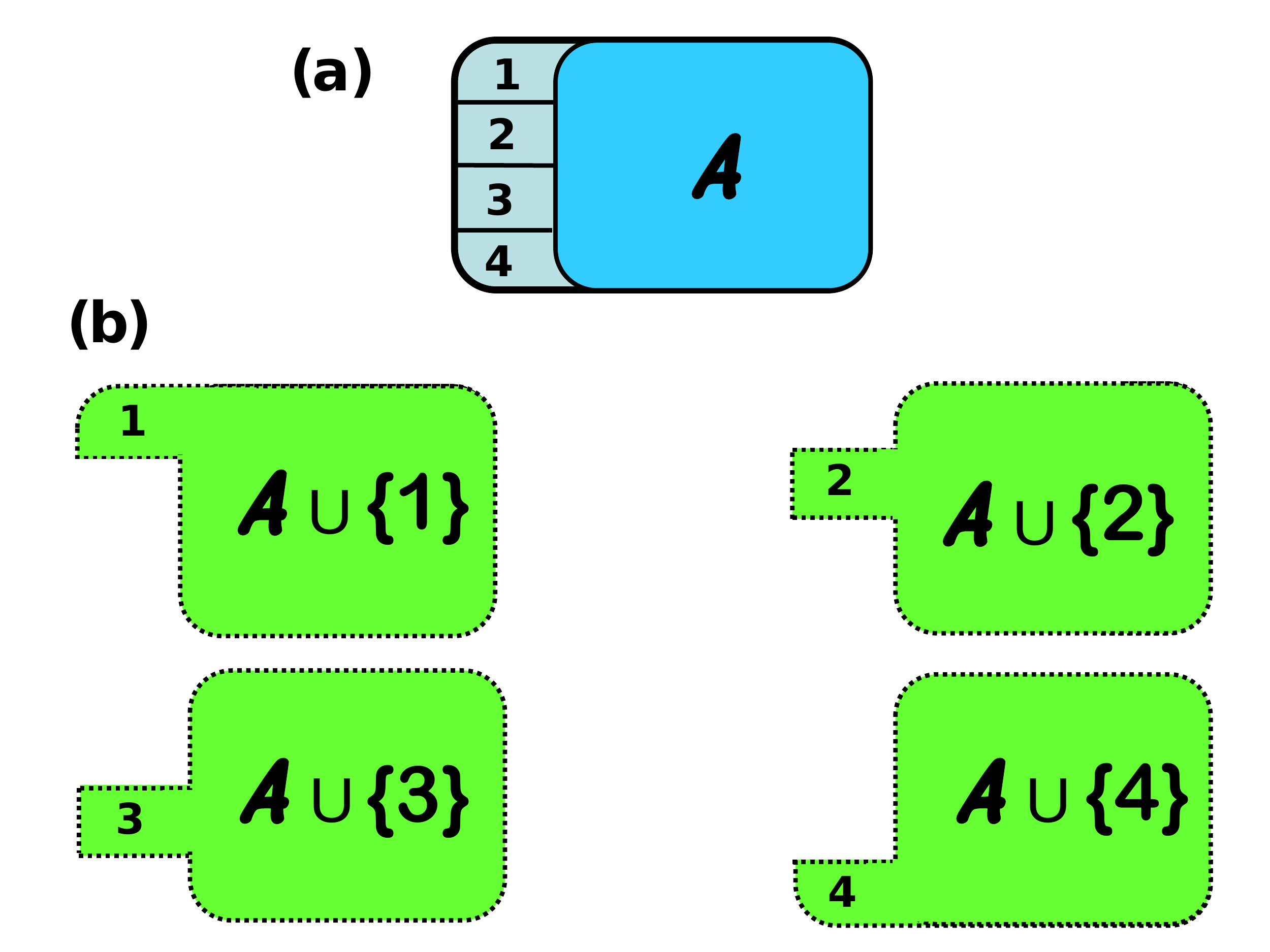}
\end{center}
\caption{An illustration of the terms in the sum in equation \eqref{eq:rhoA1234}. The set $\set{A}$, a subset of the total system $\set{S}$, is shown in blue in (a), along with four other sets $1,2,3,4$ distinct from $\set{A}$. Then in (b) in green we show the four different sets that that can be made from the union of $\set{A}$ and one of the other sets.  We have omitted the subscripts on the set variables.}
 \label{fig:EOMpart}
\end{figure}

To see how this works let us consider a simple example. Suppose one has an $N$-cell system ${\cal S}$, and we define a specific sub-set ${\cal A}_{\alpha}^{(n)}$ of ${\cal S}$ by removing 4 designated cells from ${\cal S}$ (so that $n = N-4$).  This example is illustrated in Fig. \ref{fig:EOMpart}. Writing out the sum over $\ell$ in equation \eqref{eq:rhoA} explicitly we have (omitting the subscripts on the set variables),
\begin{widetext}
\begin{equation}
 \label{eq:rhoA1234}
i\hbar\partial_t\bar\rho_{\mathcal{A}}\;\;=\;\; \left[\bar{H}_{\mathcal{A}},\bar\rho_{\mathcal{A}}\right] \;+\; \sum_{j\in\mathcal{A}}\left(\tr{1}\left[V_{j1},\bar\rho_{\mathcal{A}\cup\{1\}}\right]+
\tr{2}\left[V_{j2},\bar\rho_{\mathcal{A}\cup\{2\}}\right]+\tr{3}\left[V_{j3},
\bar\rho_{\mathcal{A}\cup\{3\}}\right]+\tr{4}\left[V_{j4},
\bar\rho_{\mathcal{A}\cup\{4\}}\right]\right).
\end{equation}
\end{widetext}
and we see explicitly how the equation of motion for the $(N-4)$-cell system ${\cal A}_{\alpha}^{(n)}$ involves a coupling between the $(N-4)$-cell density matrix $\rho_{\mathcal{A}_{\alpha}^{(n)}}$ and a set of $(N-3)$-cell density matrices $\bar\rho_{\mathcal{A}_{\alpha}^{(n)}\cup\{ \ell \}}$, with $\ell = 1,2...4$.


In the next sub-section we discuss the example of a spin system; this will allow us to work out expressions like this explicitly.

As already noted above, there is a loose analogy here with the Schwinger-Dyson equations in quantum field theory and in non-relativistic many-body theory, in that we end up with a chain of coupled integro-differential equations for the $\bar\rho_{\mathcal{A}_{\alpha}^{(n)}}$ (here we restore the indices $\alpha$ and $n$, to emphasize that we are dealing in all these equations with a specific subset of ${\cal S}$ in which $n$ denotes the number of cells involved, and $\alpha$ the specific set of $n$ cells that has been chosen).


\subsection{Hierarchy of equations for reduced density matrices N Spin$-1/2$'s.}


For our set of $N$ spin$-\tfrac{1}{2}$'s, the $\{\boldsymbol{\sigma}_i\}$, the cells again become individual sites, each with its own spin. We wish to find the dynamics of the various spin correlators, following the general theory given in section \ref{sec:EOM-site}.  For this we need a Hamiltonian for the $N$-spin system. The general pairwise interaction Hamiltonian for this case is:
\begin{equation}
H \;=\; \sum_i\tfrac{1}{2}\vc{h}_i\cdot\boldsymbol{\sigma}_i+
\sum_{i=1}^N\sum_{j<i}\tfrac{1}{2}V_{ij}^{\mu\nu}\sigma_i^\mu\sigma_j^\nu.
 \label{pairSpinH}
\end{equation}

In this Hamiltonian each spin feels a local field ${\bf h}_j$, and we have a pairwise interaction $V_{ij}^{\mu\nu}$ between the spins. Commonly used examples are (i) the quantum Ising model, for which ${\bf h}_i = h_o \hat{\bf x}$ and $V_{ij}^{\mu\nu}\sigma_i^\mu\sigma_j^\nu = V_{ij}^{zz}\delta^{\mu z} \delta^{\nu z}$, and (ii) the nearest-neighbour Heisenberg model, where ${\bf h}_i = {\bf h}$ is a uniform external field, and $V_{ij}^{\mu\nu}\sigma_i^\mu\sigma_j^\nu = J_o \delta^{\mu\nu}$, with $i,j$ restricted to be nearest neighbours.

In what follows we first derive the general hierarchy of equations of motion, and then look at some simple spacial cases.

\subsubsection{General Form of Hierarchy}
 \label{sec:spinH}

We derive the equations of motion for the various spin correlators from the reduced density matrix equation of motion we have found in \eqref{eq:rhoA}. Again, we pick a specific subset $\mathcal{A}$ of the total $N$-spin system; we will therefore be interested in the time evolution of expectation values of products of spin operators for spins in $\mathcal{A}$.

The result of the calculation can be read off from the general equation of motion in (\ref{eq:rhoA}); the commutators are evaluated in Appendix \ref{sec:App-B2}, and we find
\begin{widetext}
\begin{align}
\frac{\ud}{\ud t}\Bigl\langle\prod_{i\in\mathcal{A}}\sigma_i^{\mu_i}\Bigr\rangle \;\;\;=\;\;\;
&\sum_{i\in\mathcal{A}}\varepsilon^{\mu_i\alpha\nu}h_i^\alpha
\Bigl\langle\sigma_i^\nu\prod_{j\in\mathcal{A}\backslash\{i\}}\sigma_j^{\mu_j}
\Bigr\rangle \;\;+\;\; \sum_{i\in\mathcal{A}}\sum_{j\in\mathcal{A}\backslash \{i\}}\varepsilon^{\mu_i\alpha\nu}V_{ij}^{\alpha\mu_j}
\Bigl\langle\sigma_i^\nu\prod_{k\in\mathcal{A}\backslash\{i,j\}}\sigma_k^{\mu_k}\Bigr\rangle  \nonumber\\
& \qquad\qquad\qquad\qquad +\;\; \sum_{i\in\mathcal{A}}\sum_{\ell\not\in\mathcal{A}}
\varepsilon^{\mu_i\alpha\nu}V_{i\ell}^{\alpha\lambda}
\Bigl\langle\sigma_\ell^\lambda\sigma_i^\nu\prod_{j\in\mathcal{A}\backslash\{i\}}
\sigma_j^{\mu_j}\Bigr\rangle
 \label{eq:EOMgensp12}
\end{align}
\end{widetext}
in which we see the characteristic form of a coupled hierarchy of differential equations: the time derivative of the correlator is given in terms of correlators between spins in $\mathcal{A}$ and correlators among all possible subsets of $\mathcal{A}$ with one spin removed, as well as all possible sets made from adding one spin to $\mathcal{A}$. The local field term mixes up the different correlators between qubits in the cluster of qubits $\mathcal{A}$, while the interaction terms ``transfers correlations'' to clusters which contain either one less or one more qubit.

The result (\ref{eq:EOMgensp12}) is still rather forbidding, mainly because it describes the dynamics of correlators for {\it all} of the spins contained in $\mathcal{A}$.  To make it more transparent, we now consider two special cases of this general result.

\subsubsection{One- and two-spin Correlators}
 \label{sec:12spinC}

To simplify eqtn. (\ref{eq:EOMgensp12}), we can make the subset $\mathcal{A}$ small. We consider the two simplest cases, where $\mathcal{A}$ includes one or two sites.

\vspace{2mm}

{\bf Single-site $\mathcal{A}$}: Suppose $\mathcal{A}$ is just a single spin - without loss of generality we call this ``site 1". Then there is only one correlator, given by the expectation value $\langle \sigma_1^{\mu} (t) \rangle$; the equation of motion, read off from (\ref{eq:EOMgensp12}), is just
\begin{equation}
 \frac{\ud}{\ud t}\langle\sigma_1^{\mu}\rangle \;=\;
 \varepsilon^{\mu_1\alpha\beta} \left(h_1^{\alpha}
\langle\sigma_1^{\beta} \rangle + \sum_{\ell \neq 1} V_{1\ell}^{\alpha \lambda} \langle \sigma_{\ell}^{\lambda} \sigma_1^{\beta}\rangle\right)
 \label{1spinC}
\end{equation}
where we recall that $V_{ii}^{\alpha\beta} = 0$, ie., there is no on-site interaction apart from the local field ${\bf h}_i$, and we note again that the product over an empty set just gives unity for the 3rd term in (\ref{eq:EOMgensp12}). In vector notation eq. (\ref{1spinC}) reads
\begin{equation}
 \frac{\ud}{\ud t} \langle\boldsymbol{\sigma}_1 \rangle \;=\; ({\bf h}_1 + {\bf \tilde{V}}_1) \times \langle \boldsymbol{\sigma}_1 \rangle
 \label{1spinCV}
\end{equation}
where the total field ${\bf \tilde{V}}_1$ acting on $\boldsymbol{\sigma}_1$ from all the other spins, via the interaction, has components
\begin{equation}
\tilde{V}_1^{\alpha} \;=\; \sum_{\ell \neq 1} V_{1\ell}^{\alpha \lambda} \langle \sigma_{\ell}^{\lambda} \rangle
 \label{Via-comp}
\end{equation}
Thus (\ref{1spinCV}) is simply telling us that spin 1 is precessing in a total field coming from the local external field plus the field on site 1 generated by all the other spins, via the interaction.

This result is of course well known, and can be derived trivially starting directly from the Hamiltonian. The second term in (\ref{1spinCV}) can be thought of as a ``Hartree" mean field interaction term.

\vspace{2mm}

{\bf Two-site $\mathcal{A}$}:  Slightly less trivial is the result we get when $\mathcal{A}$ incorporates a pair of sites, which we call site 1 and site 2. We are then interested in the dynamics of the pair correlator $\langle \sigma_1^{\mu_1} \sigma_2^{\mu_2} \rangle$, and we find
\begin{eqnarray}
 \frac{\ud}{\ud t}\langle\sigma_1^{\mu_1}\sigma_2^{\mu_2}\rangle
 \;&=& \sum_{j \neq j' = 1,2} \varepsilon^{\mu_j\alpha\beta} \;
\biggl[ h_j^{\alpha} \langle \sigma_j^{\beta}
 \sigma_{j'}^{\mu_{j'}}\rangle + V_{12}^{\alpha \mu_{j'}} \langle \sigma_j^{\beta} \rangle  \nonumber \\
 &&\qquad + \; \sum_{\ell \neq 1,2} V_{j\ell}^{\alpha\lambda} \langle
 \sigma_{\ell}^{\lambda} \sigma_j^{\beta} \sigma_{j'}^{\mu_{j'}}
 \rangle
\biggr] \qquad
 \label{2spinC}
\end{eqnarray}
where $\sum_{j \neq j' = 1,2}$ means that we sum over both $j$ and $j'$, with the restriction that $j \neq j'$. This result contain both the fields we already saw for the single-site correlator (but now acting on both spins) plus a term - the 2nd term on the RHS in   (\ref{2spinC}) above - which accounts for the interaction between the two spins.

We can now see intuitively how the results will develop as one goes to correlators including larger numbers of spins in $\mathcal{A}$. It is also interesting to see how things simplify if we look at a very small total system. Thus, eg., suppose system ${\cal S}$ comprises only $N=2$ spins. Then the sub-system $\mathcal{A}$ is just the whole system, and we expect the result to be trivial. Writing out all terms explicitly, we have
\begin{eqnarray}
 \frac{\ud}{\ud t}\langle\sigma_1^{\mu_1}\sigma_2^{\mu_2}\rangle
 \;&=&  \varepsilon^{\mu_1\alpha\beta} (
h_1^{\alpha} \langle \sigma_1^{\beta}
 \sigma_{2}^{\mu_{2}}\rangle + V_o^{\alpha \mu_{2}} \langle \sigma_1^{\beta} \rangle)  \nonumber \\
 && + \; \varepsilon^{\mu_2\alpha\beta}
(h_2^{\alpha} \langle \sigma_2^{\beta}
 \sigma_{1}^{\mu_{1}}\rangle  + V_o^{\alpha \mu_{1}} \langle \sigma_2^{\beta} \rangle) \qquad
 \label{2spin2}
\end{eqnarray}
where we have written $V_{12} = V_o$ for the interspin interaction; the role of the effective fields acting on the one- and two-spin correlators is now transparent.

\subsubsection{Relationship to Schwinger-Dyson Hierarchy}
 \label{sec:SchwD}

The Schwinger-Dyson hierarchy \cite{schwingerD,martinS} exists in both relativistic and non-relativistic forms - it is an infinite chain of coupled equations of motion for $n$-point correlators, whose specific form depends on the interactions in the theory being treated. Its general form is similar to the classical BBGKY hierarchy \cite{BBGKY}.

To see how this related to what we have done, consider the Schwinger-Dyson hierarchy for a simple scalar field Lagrangian of form
\begin{equation}
\mathcal{L} \;\;=\;\; \tfrac{1}{2}\phi\hat{K}_{o}^{-1}\phi\;-\; V\left(\phi\right)\label{L-phi4}
\end{equation}
where $\hat{K}_{o}$ is the free field propagator. Here $x$ is a spacetime coordinate; and to be definite let us assume a simple local ``pairwise" interaction, of form
\begin{equation}
V(\phi) \;=\; {g \over 4!} \phi^4(x)
 \label{phi4}
\end{equation}

Then the Schwinger-Dyson hierarchy for the $n$-point correlation functions $G_n( \{ x_j \})$, with $j = 1, \cdots n$, is given by
\begin{widetext}
\begin{equation}
K_o^{-1}(x,x) G_{n}\left(x,x_{1}^{\prime},\dots,x_{n-1}^{\prime}\right)-
\frac{g}{6}G_{n+2}\left(x,x,x,x_{1}^{\prime},\dots,x_{n-1}^{\prime}\right)\;=\; i\hbar\sum_{j=1}^{n-1}\delta\left(x-x_{j}^{\prime}\right)\tilde{G}_{n-2}
\left(\left\{ x_{j}^{\prime}\right\} \right)
 \label{SD-phi4}
\end{equation}
where $K_o^{-1}(x,x') = \langle x |\hat{K}_o^{-1}|x' \rangle$. If we multiply (\ref{SD-phi4}) through by $\hat{K}_o$, we have
\begin{equation}
G_{n}\left(x,x_{1}^{\prime},\cdots,x_{n-1}^{\prime}\right)+
\frac{g}{6}\int d^{4}z\,K_o\left(x-z\right)G_{n+2}\left(zzz,x_{1}
^{\prime},\cdots,x_{n-1}^{\prime}\right)+i\hbar\sum_{j=1}^{n-1}K_o
\left(x-x_{j}^{\prime}\right)\tilde{G}_{n-2}\left(\left\{ x_{j}
^{\prime}\right\} \right) \;=\; 0 \;\;\;
 \label{SD-phi4'}
\end{equation}
\end{widetext}

In both of these equations we define the ``reduced" correlator
$\tilde{G}_{n-2}\left(\left\{ x_{j}^{\prime}\right\} \right)$ by
\begin{equation}
\tilde{G}_{n-2}\left(\left\{ x_{j}\right\} \right)=G_{n-1}\left(x_{1}^{\prime},\cdots x_{j-1}^{\prime},x_{j+1}^{\prime},\cdots,x_{n-1}^{\prime}\right) \;\;\;\;\;
 \label{G-tilde}
\end{equation}
from which the external legs with coordinates $x_{j}^{\prime}$ and $x_{n}^{\prime}$ have been removed.

The hierarchical form of eqtn. (\ref{SD-phi4'}), in which correlators $G_n$ are coupled to both higher and lower correlators, is very clear. Physically, one describes this equation by saying that if we have an excitation propagating from $x_j^{\prime}$ to $x$ in the presence of a set of mutually interacting excitations propagating between the points $x_{1}^{\prime},\cdots x_{j-1}^{\prime},x_{j+1}^{\prime},\cdots,x_{n-1}^{\prime}$, then it can do so with or without interacting with the other excitations.

Mathematically, we see that the main differences between the Schwinger-Dyson hierarchy and the one we have derived here are:

(i) Here we are not dealing with the propagation of correlators like $G_n$ between different spacetime intervals, but instead with time-local correlators in which space does not appear (in its place we have cell or site indices $i,j, \cdots$).

(ii) In contrast with field theory where equations simplify because the variables are indistinguishable, the variables considered here on different cells or sites are distinguishable, and each such variable has to be identified explicitly in the equations of motion. This makes the equations more complex.

One can of course integrate equations like (\ref{eq:rhoA}), (\ref{eq:rhoA1234}), or (\ref{eq:EOMgensp12}), over time - in analogy with the passage from (\ref{SD-phi4}) to (\ref{SD-phi4'}). The resulting form can be seen by choosing simple examples, such as the spin examples in (\ref{1spinCV}) or (\ref{2spinC}). The same interpretation applies - the spins in $\mathcal{A}$ can evolve with or without interacting with other spins outside $\mathcal{A}$.

Such an approach is very useful when dealing with regular lattices of, eg., spins; then we can apply decoupling techniques to the resulting hierarchy very similar to those used for the Schwinger-Dyson equations. An example appears in Gomez-Leon et al., applied to the quantum Ising model \cite{alvi18}.

However, in dealing with the general case, we would like to develop other approaches, to which we now turn.

\section{Entanglement Correlators}
\label{sec:PhysQ-B}


Although the hierarchy of equations governing the dynamics of the different density matrices has a clear physical interpretation, and allows us to formulate the idea of different levels of entanglement, the equations of motion in the form given are not all that convenient to solve.

In what follows we set up a more useful description. The basic idea is fairly simple - we define a ``supervector" whose components are an ordered list of all the different time-dependent correlation functions. We then derive a linear first-order differential equation for the time dependence of this vector. In keeping with the rest of this paper, we do not attempt to solve this equation - this will be done elsewhere, in studies of specific models.

To more easily explain the development, we do things first for a simple 2-spin problem, and then discuss some aspects of a general formulation of this kind - in particular, we describe how one treats a pair of coupled systems, and how to treat the equation of motion perturbatively, when there is a small parameter.


\subsection{Example: Entanglement correlator dynamics: two qubits}
\label{sec:2QB-Dyn}

For an arbitrary quantum system, the set of all possible observables is usually rather complicated. However in the case of spin systems, one can make an exhaustive list. For a single spin $\boldsymbol{\tau}$, the spin dynamics is completely defined by giving, as a function of time, the expectation values of all 3 components $\langle \tau_{\mu}(t) \rangle$. For a pair of spins $\boldsymbol{\tau}_1$ and $\boldsymbol{\tau}_2$, 15 different correlators are required, viz., $\langle\boldsymbol{\tau}_1(t)\rangle$, $\langle\boldsymbol{\tau}_2(t)\rangle$, and
$\langle\boldsymbol{\tau}_1\otimes\boldsymbol{\tau}_2\rangle$, where this last contains components $\langle\tau_1^{\mu}(t) \tau_2^{\nu}(t)\rangle$. For a set of $N$ qubits, we need $2^{2N}-1$ correlators.

To see how the general idea works, we go back to the the example of two qubits, with Hamiltonian
\begin{equation}
H=\sum_{a=1}^2\tfrac{1}{2}{\bf h}_a \cdot \boldsymbol{\tau}_a \;+\; \tfrac{1}{2}V_{\mu\nu}\tau_1^\mu\tau_2^\nu
 \label{2QB-ex}
\end{equation}
in which the orientation of the 2 static fields ${\bf h}_1$, ${\bf h}_2$ is arbitrary. This is just the Hamiltonian (\ref{pairSpinH}), for a pair of spins. 

Now, suppose we arrange the all the information contained in the 2-qubit density matrix (compare equation \eqref{eq:rho12}) in the form of a 15-component ``supervector" $\underline{X}$ in the ``space of possible correlators'', according to
\begin{equation} \label{eq:X2spins}
\underline{X}=\begin{pmatrix} X_1\\ X_2\\ X_3\\ X_5\\ X_6\\ X_7\\ X_8\\ \vdots\end{pmatrix}=\begin{pmatrix}\langle\tau_1^x\rangle\\
\langle\tau_1^y\rangle\\\langle\tau_1^z\rangle\\\langle
\tau_2^x\rangle\\\langle\tau_2^y\rangle\\\langle\tau_2^z\rangle\\
\langle\tau_1^x\tau_2^x\rangle\\\langle\tau_1^x\tau_2^y\rangle
\\\vdots\end{pmatrix} \equiv \begin{pmatrix}
\langle\boldsymbol{\tau}_1\rangle\\
\langle\boldsymbol{\tau}_2\rangle\\
\langle\boldsymbol{\tau}_1\otimes\boldsymbol{\tau}_2\rangle\\
\end{pmatrix}.
\end{equation}

We can then rewrite the hierarchy of equations of motion for the 2-qubit density matrix in the form
\begin{equation}
\frac{\ud}{\ud t}\underline{X} \;=\; \mathbb{M} \underline{X}
\end{equation}
or, written out explicitly, in the block structure
\begin{align}
\frac{\ud}{\ud t}\begin{pmatrix}
\langle\boldsymbol{\tau}_1\rangle\\
\langle\boldsymbol{\tau}_2\rangle\\
\langle\boldsymbol{\tau}_1\otimes\boldsymbol{\tau}_2\rangle\\
\end{pmatrix}=&\begin{pmatrix}
\mathbb{L}_1&0&\mathbb{U}_{1,p}\\
0&\mathbb{L}_2&\mathbb{U}_{2,p}\\
\mathbb{U}_{p,1}&\mathbb{U}_{p,2}&\mathbb{L}_p\\
\end{pmatrix}\begin{pmatrix}
\langle\boldsymbol{\tau}_1\rangle\\
\langle\boldsymbol{\tau}_2\rangle\\
\langle\boldsymbol{\tau}_1\otimes\boldsymbol{\tau}_2\rangle\\
\end{pmatrix}.\label{e1:hier12block}
\end{align}

Looking first at the diagonal matrix elements of $\mathbb{M}$, we see that $\mathbb{L}_1$, $\mathbb{L}_2$ are $3\times 3$ matrices which give an infinitesimal rotation of $\langle\boldsymbol{\tau}_1\rangle,\langle\boldsymbol{\tau}_2\rangle$ around the applied fields. The $9\times9$ matrix $\mathbb{L}_p$ rotates the pair correlator $\langle\boldsymbol{\tau}_1\boldsymbol{\tau}_2\rangle$ around the applied fields, and can also be written as a rank $4$ tensor (the lowered indices in the following expressions are understood to be contracted to the right in equation \eqref{e1:hier12block}). Thus we have:
\begin{align}
 \label{eq:hs}
\mathbb{L}_1^{\mu}{}_{\nu}=&h_1^\lambda\varepsilon^{\mu}{}_{\lambda\nu}\\
\mathbb{L}_2^{\alpha}{}_{\beta}=&h_2^\gamma\varepsilon^{\alpha}{}_{\gamma\beta}\\
\mathbb{L}_p^{\mu\alpha}{}_{\nu\beta}=&
\mathbb{L}_1^{\mu}{}_{\nu}\delta^\alpha{}_\beta+
\delta^\mu{}_\nu\mathbb{L}_2^{\alpha}{}_{\beta}.
\end{align}
Turning now to the non-diagonal interaction matrices, we have  terms $\mathbb{U}_{1,p},\mathbb{U}_{2,p}$ which are $3\times 9$ matrices, and which create single qubit coherences from the pair correlator; the corresponding terms  $\mathbb{U}_{p,1},\mathbb{U}_{p,2}$ are $9\times 3$ matrices which create pair coherences from the single qubit coherences. All of these interaction matrices  may be represented as rank 3 tensors:
\begin{align}
 \label{eq:us}
\mathbb{U}_{1,p}{}^\mu{}_{\nu\beta}=&V_{\lambda\beta}\varepsilon^{\mu\lambda}{}_\nu\\
\mathbb{U}_{p,1}{}^{\mu\alpha}{}_{\nu}=&V_{\lambda}{}^{\alpha}\varepsilon^{\mu\lambda}{}_\nu\\
\mathbb{U}_{2,p}{}^\alpha{}_{\nu\beta}=&V_{\nu\gamma}\varepsilon^{\alpha\gamma}{}_\beta\\
\mathbb{U}_{p,2}{}^{\mu\alpha}{}_{\beta}=&V^\mu{}_{\gamma}\varepsilon^{\alpha\gamma}{}_\beta.
\end{align}

We see that the matrix $\mathbb{M}$ is fully anti-symmetric and has eigenvalues which are either zero or pure imaginary. One can divide these into two classes, as follows:

(i) there are at least 3 zero eigenvectors of  $\mathbb{M}$, which are linear combinations of the eigenstates $|n\rangle\langle n|$ of the Hamiltonian (the dimensionality of the system of equations is one less than the number of components of the density matrix, because the equations automatically preserve the trace of the density matrix). 

(ii) The other eigenvalues of $\mathbb{M}$ occur at every difference $E_n-E_m$ in the eigenvalues of the Hamiltonian and their eigenvectors are off-diagonal elements of the density matrix  $|n\rangle\langle m|$. 

In general we can define a set of Green functions $\{\mathbf{g}_{ij}\}$ with $i,j\in\{1,2,p\}$ for the solution to the equations of motion \eqref{e1:hier12block}, so that the solution to the equations of motion for the vector $\underline{X}$ can be written
\begin{equation}
\underline{X}(t)=\mathbb{G}(t)\underline{X}(0)
\end{equation}
where the total propagator has the block form
\begin{align}
\mathbb{G}(t) \;\;=\;\;
\begin{pmatrix}
\vc{g}_{11}(t)&\vc{g}_{12}(t)&\vc{g}_{1p}(t)\\
\vc{g}_{21}(t)&\vc{g}_{22}(t)&\vc{g}_{2p}(t)\\
\vc{g}_{p1}(t)&\vc{g}_{p2}(t)&\vc{g}_{pp}(t)\\
\end{pmatrix}
\label{e1:gf12block}
\end{align}
A formal solution for this Green function is found by Laplace transforming; writing $f(z)=\int^\infty_{0}\text{d}t\,f(t)e^{-zt}$, we have
\begin{equation}
\mathbb{G}(z)\;=\;\left[z\mathbb{I}-\mathbb{M}\right]^{-1}.
\end{equation}
so that $\mathbb{G}(z)$ has poles at along the imaginary axis at all the differences between the energy eigenvalues $\pm i\Delta E$ as well as a pole at zero with a degeneracy of at least four.

\begin{figure}
\begin{center}
\includegraphics[width=8.5cm]{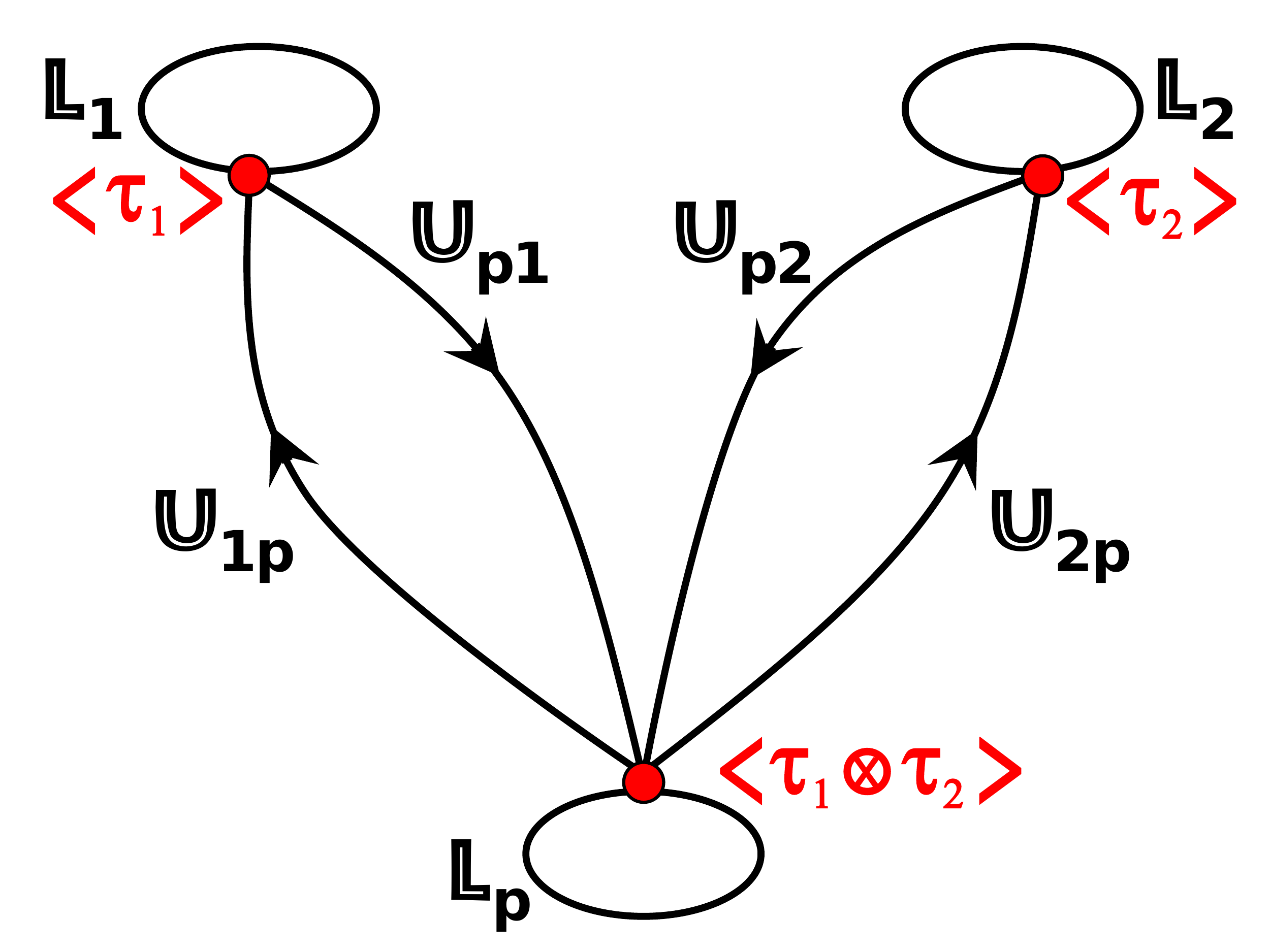}
\end{center}
\caption{Graphical representation of a term in the expansion of $\mathbb{G}(t)$ as an exponential power series (cf. eqtn. (\ref{eq:gf12t})). In this term all the entries in the matrix $\mathbb{M}$ appear (these entries are given in eqtn. (\ref{e1:hier12block})). The correlators $\langle\boldsymbol{\tau}_1\rangle$, $\langle\boldsymbol{\tau}_2\rangle$, and $\langle\boldsymbol{\tau}_1\otimes\boldsymbol{\tau}_2\rangle$ are shown as red vertices, the interaction matrices $\mathbb{U}_{1,p}$, $\mathbb{U}_{2,p}$, $\mathbb{U}_{p,1}$, and $\mathbb{U}_{p,2}$ are shown as directed lines, and the rotation matrices $\mathbb{L}_1$, $\mathbb{L}_2$, and $\mathbb{L}_p$ are shown as undirected lines.        }
 \label{fig:expM}
\end{figure}

In the time domain the Green function is just
\begin{equation}
\mathbb{G}(t)=\exp\left(\mathbb{M}t\right)=\sum_{n=0}^\infty \frac{\mathbb{M}^nt^n}{n!}.\label{eq:gf12t}
\end{equation}
This series can be represented graphically (see Fig. \ref{fig:expM}). We define a graph whose
vertices are the possible correlators, having (directed) links between them which
represent the block components of  $\mathbb{M}$. Then we an $n$-th order term in
the sum is represented by a ``walk" (ie., sequence of $n$ hops) across $n$ links
between nodes; multiplying each term by $t^n/n!$ we get the Green function.


It is important to get an idea of what these expressions look like in practice. Suppose we look first at a very simple case, where the Hamiltonian is
\begin{equation}
H=\tfrac{1}{2} \left[ \Delta_1 \tau^x+\Delta_2\sigma^x+\omega\tau^z\sigma^z \right]
 \label{eq:HDeltaomega2}
\end{equation}
having energy eigenvalues $\pm\epsilon_1,\pm\epsilon_2$ with 
\begin{align}
\epsilon_1&=\frac{1}{2}\sqrt{\omega^2+(\Delta_1+\Delta_2)^2}\\
\epsilon_2&=\frac{1}{2}\sqrt{\omega^2+(\Delta_1-\Delta_2)^2}.
\end{align}

The 225 elements in the $15 \times 15$ matrix $\mathbb{M}$ can now be written out directly, using eqtns. (\ref{eq:hs}) - (\ref{eq:us}). The large majority of the elements are zero; the non zero eigenvalues of $\mathbb{M}$ for this case are 
\begin{align}
\omega_{10}&=(\epsilon_1-\epsilon_2)\\
\omega_{20}&=\epsilon_1+\epsilon_2\\
\omega_{30}&=2\epsilon_1\\
\omega_{21}&=2\epsilon_2.
\end{align}

The different components of $\mathbb{G}(t)$, ie., the 9 different matrix Green functions given in eqtn. (\ref{e1:gf12block}) are then multiperiodic functions containing these 4 frequencies. Their explicit expressions are of course quite lengthy to write out; in App. \ref{sec:App-C} the explicit results for $\mathbb{G}(z)$ are written in full. 

The general 2-qubit Hamiltonian (\ref{2QB-ex}) is not much more complicated than this. In particular, the matrix $\mathbb{M}$ has the key property that it is rather sparse, ie., most elements are still zero. To see this, we write the interaction tensor in diagonal form, ie., $V_{\mu\alpha}=V_{xx}\hat{x}_\mu\hat{x}_\alpha+V_{yy}\hat{y}_\mu\hat{y}_\alpha+V_{zz}\hat{z}_\mu\hat{z}_\alpha$;  note that there is always a co-ordinate system where $V_{\mu\alpha}$ is of this form, which can be obtained using the singular value decomposition of $V_{\mu\alpha}$. Then the sub-matrices which make up $\mathbb{M}$ can be written explicitly as
\begin{align}
\mathbb{L}_1=&\begin{pmatrix}
0&-h_1^z&h_1^y\\
 h_1^z &0&-h_1^z\\
 -h_1^y&h_1^z&0
\end{pmatrix}\\
\mathbb{L}_2=&\begin{pmatrix}
0&-h_2^z&h_2^y\\
 h_2^z &0&-h_2^z\\
 -h_2^y&h_2^z&0
\end{pmatrix}\\
\mathbb{L}_p=&\begin{pmatrix}
0& -h_2^z & h_2^y & -h_1^z &0&0& h_1^y &0&0\\
h_2^z & 0 & -h_2^x & 0 & -h_1^z &0&0& h_1^y &0 \\
-h_2^y & h_2^x &0&0&0& -h_1^z &0&0& h_1^y\\
h_1^z &0&0&0& -h_2^z & h_2^y & -h_1^x &0&0\\
0& h_1^z &0& h_2^z &0& -h_2^x &0& -h_1^x &0\\
0&0& h_1^z & -h_2^y & h_2^x &0&0&0& -h_1^x\\
-h_1^y &0&0& h_1^x &0&0&0& -h_2^z & h_2^y\\
0& -h_1^y &0&0& h_1^x &0& h_2^z &0& -h_2^y\\
0&0& -h_1^y &0&0& h_1^x & -h_2^y & h_2^x & 0
\end{pmatrix}\\
\mathbb{U}_{1p}=&\begin{pmatrix}
0&0&0&0&0& -V_{zz} &0& V_{yy} &0\\
0&0& V_{zz} &0&0&0& -V_{xx} &0&0\\
0& -V_{yy} &0& V_{xx} &0&0&0&0&0
\end{pmatrix}\\
\mathbb{U}_{2p}=&\begin{pmatrix}
0&0&0&0&0& V_{yy} &0& -V_{zz} &0\\
0&0& -V_{xx} &0&0&0& V_{zz} &0&0\\
0& V_{xx} &0& -V_{yy} &0&0&0&0&0
\end{pmatrix}\\
\mathbb{U}_{p1}=&-\mathbb{U}_{1p}^T\\
\mathbb{U}_{p2}=&-\mathbb{U}_{2p}^T.
\end{align}
By making ${\bf h}_a = \hat{x} \Delta_a$, and using a purely longitudinal coupling, we get back the simpler Hamiltonian in (\ref{eq:HDeltaomega2}). In any case, we see that most elements in these matrices are zeroes. 

More generally, so long as there are only local fields and pairwise interactions, it is evedent that the ``sparseness'' of the matrix $\mathbb{M}$ will increase rapidly with the number of qubits. We will see in future papers that this makes them very useful in practical problems involving many interacting qubits.


\subsection{Remarks on a General Formulation}
\label{sec:EOM-gen}

Let us now consider how this might go for more complicated systems. The generalization of the 2-spin results to $N$ spins is clear - now the supervector $\underline{X}$ has $d_N = 2^{2N} - 1$ entries, growing very rapidly with $N$.

More generally one may have to deal with systems in the thermodynamic limit, having an infinite number of degrees of freedom. Moreover, most degrees of freedom in Nature are usually described by continuous variables, and this automatically leads to an infinite set of possible correlators (such as the set $\langle x \rangle, \langle x, x' \rangle, \langle x, x', x'' \rangle, \cdots$, etc, for a single coordinate degree of freedom); and as noted in the introduction, the system may be composed of indistinguishable particles.

We will not deal with all these complications here - but it is still useful to understand some more general features of problems involving distinguishable spins. In what follows we look at two key questions, viz., (i) how do things work when we have 2 coupled spin systems, and (ii) if there is a small parameter in the problem, how do we make perturbation expansions for the entanglement correlators?

\subsubsection{Two Coupled Systems}
\label{sec: 2C-Sys}

The special case of two separate but coupled systems is of interest for several reasons. Most notably, it forms the basis for a discussion of a central system coupled to some environment; and it is also useful when one comes to analyze how entanglement develops between any pair of systems.

Consider a pair of systems ${\cal S}_1$ and ${\cal S}_2$, which may or may not interact, and which are in general entangled.
We again define an abstract supervector $\underline{\xi}$ which contains all possible correlators for the pair of systems, in the form
\begin{equation}
\underline{\xi} \;=\; \begin{pmatrix}
\underline{X}_1\\
\underline{Y}\\
\underline{X}_2\\
\end{pmatrix}\label{eq:xidef}
\end{equation}
where $\underline{X}_1$ is the vector containing all the correlators of operators acting on ${\cal S}_1$ alone, $\underline{X}_2$ likewise for ${\cal S}_2$, and $\underline{Y}$ refers to all ``joint" operators, acting on both systems together.

As an example one can consider a pair of qubit systems, one containing $n_1$ spins $\{ \boldsymbol{\tau}_i \}$, and the other $n_2$ spin-$1/2$ degrees of freedom $\{ \boldsymbol{\sigma}_j \}$, with the total number of spins being $N = n_1 + n_2$. We then have
\begin{equation}
\underline{X}_1=\begin{pmatrix}\langle\tau_1^\alpha\rangle
\\\langle\tau_2^\alpha\rangle\\\vdots\\\langle\tau_1^\alpha\tau_2^\beta\rangle
\\\langle\tau_1^\alpha\tau_3^\beta\rangle\\\vdots\\
\langle\tau_1^\alpha\tau_2^\beta\tau_3^\delta\rangle\\\vdots\end{pmatrix} \;\;\;\; ; \qquad
\underline{X}_2=\begin{pmatrix}\langle\sigma_1^\alpha\rangle
\\\langle\sigma_2^\alpha\rangle\\\vdots\\
\langle\sigma_1^\alpha\sigma_2^\beta\rangle\\
\langle\sigma_1^\alpha\sigma_3^\beta\rangle\\\vdots\\
\langle\sigma_1^\alpha\sigma_2^\beta\sigma_3^\delta\rangle\\\vdots\end{pmatrix}
\end{equation}
for the supervectors of ${\cal S}_1$ and ${\cal S}_2$ respectively; the supervector $\underline{Y}$ on the other hand has the entries
\begin{equation}
\underline{Y}=\begin{pmatrix}\langle\tau_1^\mu\sigma_1^\alpha\rangle
\\\langle\tau_1^\mu\sigma_2^\alpha\rangle\\\vdots\\
\langle\tau_2^\mu\sigma_1^\alpha\rangle\\\langle\tau_2^\mu\sigma_2^\alpha\rangle
\\\vdots\\\langle\tau_1^\mu\tau_2^\nu\sigma_1^\alpha\rangle\\\vdots\\
\langle\tau_1^\mu\sigma_1^\alpha\sigma_2^\beta\rangle\\\vdots\end{pmatrix}
\end{equation}
The number of components of these different vectors are then given by
\begin{eqnarray}
d_{X_1} &=& 2^{2n_1} - 1   \qquad\qquad d_{X_2} = 2^{2n_2} - 1 \nonumber\\
d_{Y} &=& (2^{2n_1} - 1) (2^{2n_2} - 1)  \nonumber \\
d_{X_1 + X_2} &\equiv& d_N \;=\; 2^{2N} - 1
\end{eqnarray}
where $ N = (n_1 + n_2)$.

Let us now take the Laplace transform of $\underline{\xi}(t)$, defined as before by
\begin{equation}
\underline{\xi}(z)=\int_0^\infty\ud t e^{-zt}\underline{\xi}(t)
\end{equation}
The equations of motion can then be written in the following form
\begin{equation}
\underline{\xi}(z)=\underline{\mathbb{G}}(z)\underline{\xi}(0)
\end{equation}
where $\underline{\xi}(0)$ is the initial value of $\underline{\xi}$  and $\mathbb{G}(z)$ is a matrix, whose inverse has the following block structure:
\begin{align}
\underline{\mathbb{G}}(z)^{-1}
&
=\begin{pmatrix}
\vc{g}_{1}^{-1}(z)&-\mathbb{V}_{1\set{M}}&0\\
-\mathbb{V}_{\set{M}1}&\;\;\; \vc{g}^{-1}_{\set{M}}(z)-\mathbb{V}_{\set{M}\set{M}}  \;\;\;&-\mathbb{V}_{\set{M} 2}\\
0&-\mathbb{V}_{2 \set{M}}&\vc{g}_{2}^{-1}(z)
\end{pmatrix}
\end{align}
where the ``mixed" propagator $\vc{g}_{\set{M}}$ in the middle matrix element $\mathbb{M}_{\set{M}\,\set{M}}$ is given by
\begin{equation}
\vc{g}^{-1}_{\set{M}}(z) \;\;=\;\; \vc{g}_{1}^{-1}(z)\mathbb{I}_2 + \vc{g}_{2}^{-1}(z)\mathbb{I}_1 - z \mathbb{I}_1 \mathbb{I}_2
 \label{MMM}
\end{equation}

In these equations $\underline{X}_j(z)=\vc{g}_{j}(z)\underline{X}(0)$ is the solution to the equations of motion for the individual system $j$ (with $j = 1,2$) in the absence of any coupling between them; $\mathbb{I}_{j}$ is the identity acting on system $j$, and the interaction matrix $\underline{\mathbb{V}}$ has the form, in the same $d_N$-dimensional space,
\begin{equation}
\underline{\mathbb{V}}=\begin{pmatrix}
0&\mathbb{V}_{1\set{M}}&0\\
\mathbb{V}_{\set{M}1}& \;\;\mathbb{V}_{\set{M}\set{M}} \;\;&\mathbb{V}_{\set{M}2}\\
0&\mathbb{V}_{2\set{M}}&0
\end{pmatrix}.
\end{equation}

The elements of the sub matrices of $\mathbb{V}$ can be obtained as needed by reading them off from the equations of motion (for which of course we require a specific Hamiltonian).

In general the $\underline{\mathbb{G}}(z)$ will have poles at $z=i(\omega_n-\omega_m)$ for all $n,m$  where $\omega_n$  is the $n$th energy eigenvalue of the Hamiltonian. The pole at $z=0$ will be of at least of order $2^{n_1-n_2}-1$, with larger orders occurring when the system has degenerate energy levels.

When the systems are large it does not make sense to be enumerating all the poles and their residues. Instead we simply define a spectral function which gives us the density of the poles along the imaginary axis; we write
\begin{equation}
\mathbb{A}(\omega) = {1 \over 2\pi} [\mathbb{G}(i\omega + \epsilon) - \mathbb{G}(i\omega - \epsilon)] 
 \label{specA}
\end{equation}
where we choose $\epsilon$ to be small but still larger than the typical separation between poles. For sufficiently large systems the poles will become so close that we can treat them as defining a branch cut along the imaginary axis, with magnitude $\mathbb{A}(\omega)$.

\subsubsection{Perturbation Expansions}
\label{sec:pertEx}

Suppose we have solved the full hierarchy in some specific case, and we add a small term to the Hamiltonian, - this could be, eg., to each of the bath spin local fields, or to the interaction between the central systems and the bath spins. The question is how a perturbation theory will be structured.

We do not give a full treatment here, since it is rather messy. The simplest case is the one in which we treat the interaction term $\underline{V}$ as a perturbation. We can then write an equation for the full Green function, $\underline{\mathbb{G}}(z)$ as an expansion about the $\underline{\mathbb{V}}=0$ Green function, $\underline{\mathbb{G}}_0(z)$, where in this case one has
\begin{equation}
\underline{\mathbb{G}}_0(z)\equiv\begin{pmatrix}
\vc{g}_1(z)&0                   &0\\
0          &\vc{g}_{\set{M}}(z)& 0\\
0          &		0			& \vc{g}_{2}(z)
\end{pmatrix}.
\end{equation}

A Dyson series for $\underline{\mathbb{G}}(z)$ may then be obtained in through the usual manipulations,
\begin{align}
\underline{\mathbb{G}}(z)=
&\left[\underline{\mathbb{G}}_0^{-1}(z)-\underline{\mathbb{V}}\right]^{-1}\\
=&\underline{\mathbb{G}}_0(z)\sum_{n=0}^\infty
\left(\underline{\mathbb{V}}\underline{\mathbb{G}}_0(z)\right)^{n}
\end{align}
where the matrix being raised to the $n$-th power is just
\begin{align}
\underline{\mathbb{V}}\underline{\mathbb{G}}_0(z) \;=\;   \begin{pmatrix}
0							     & \mathbb{V}_{1\set{M}}\vc{g}_\set{M}(z) &0\\
\mathbb{V}_{\set{M}1}\vc{g}_1(z) & \mathbb{V}_{\set{MM}}\vc{g}_\set{M}(z) & \mathbb{V}_{\set{M}2}\vc{g}_2(z) \\
0								&\mathbb{V}_{2\set{M}}\vc{g}_\set{M}(z) &0
\end{pmatrix}
\end{align}

Note that care needs to be taken when this expansion is performed near the high order poles of $\underline{\mathbb{G}}_0(z)$, to ensure that the corrections are still small.


\section{Summary}
 \label{sec:summary}


For the most part the results in this paper have been rather formal. Our main goal was to derive a closed set of equations of motion for the partitioned density matrices, and from there derive
coupled equations of motion for all the different correlation functions that exist for the system. This we have done in this paper, for the case of non-relativistic $N$-body quantum systems with distinguishable degrees of freedom.

Application of this analysis to the particular case of interacting qubits brings out a number of interesting features. We see clearly that the use of the entanglement correlators is in many ways a more transparent way of characterizing multipartite entanglement than the entanglement measures that have been discussed in the literature. When things are rewritten in terms of supervectors of entanglement correlators, one finds that that the resulting matrix equations of motion involve sparse matrices, which clearly makes them practically useful. 

Clearly the demonstration of the utility of these equations will come in their application to real physical systems, and this requires solutions to the equations of motion. As is always the case, such solutions require approximation techniques; in the paper we simply sketched how perturbation expansions work, leaving aside the main approximation techniques for future papers, since they need to be developed for specific models.

As already noted in the introduction, our interest in carrying out this work was partly motivated by a desire to understand how multipartite entanglement and its dynamics can be formulated for $N$-body systems, including quantum information processing systems. Our results can be applied immediately to treat the dynamics of spin systems, and elsewhere we have done this for the quantum Ising model \cite{alvi18}. These results can be applied directly to a variety of quantum magnetic systems, to spins in semiconductors, and to ions interacting in ion traps.

However the most interesting application of the techniques and results developed here may be to the dynamics of both entanglement and decoherence in systems which are coupled to an environment. A key goal of future work will be to use this work to analyze such problems. One very useful model developed for this purpose is the ``central spin" model \cite{Cspin}, in which a qubit couples to a spin bath, and generalizations of it in which the central system comprises many qubits, or is some other sort of central system. The work done here can be adapted very simply to these models.

Another useful model is the ``spin-boson" model \cite{spinB}, where a central qubit couples to an oscillator bath (with analogous generalizations to other kinds of central system). To deal with models like this we need to adapt the work done here to systems of indistinguishable degrees of freedom.

Finally, one can generalize this work to relativistic quantum fields (which of course involves indistinguishable field excitations). A scheme for this has been developed recently \cite{jordan18}, and applied to the problem of soft photon and soft graviton emission in linearized quantum gravity, where it is relevant to the black hole information problem, and to information loss during scattering processes between interacting quantum fields.


\section{Acknowledgements}
 \label{sec:ack}


We would like to thank Dr. A Gomez-Leon for extensive discussions of this work while it was in progress. We also thank Drs. A Morello, JM Raimond, and M Troyer for useful remarks. The work was supported by the National Scientific and Engineering Research Council of Canada.

\appendix


\section{Properties of Entanglement Density Matrices}
 \label{sec:App-A}


In this Appendix we prove two properties of the entanglement density matrices that were quoted without proof in section \ref{sec:rho+C-3}. We use same notation as that defined in this section.


\subsection{Proof of Eqtn. \eqref{eq:rhoC}}
 \label{sec:App-A1}


We wish here to prove the result given in eqtn. \eqref{eq:rhoC-1} (or, equivalently eq. \eqref{eq:rhoC}) for the entanglement correlated density matrices.

We do this by induction. The $n=2$ case comes from tracing out all of $\set{S}$ except $i$ and $j$ from the equation for the density matrix \eqref{eq:rhofull}, so that
\begin{equation}
\bar{\rho}_{ij}\;=\; \bar{\rho}_i\bar{\rho}_j+\bar{\rho}_{ij}^C\,\;\;\Rightarrow\,\;\; \bar{\rho}_{ij}^C \;=\;
\bar{\rho}_{ij}-\bar{\rho}_i\bar{\rho}_j
\end{equation}
as required. Now we make the inductive assumption that for all $k<n$ and $\set{B}_k\subset\set{A}_n$ we have
\begin{align}\label{eq:rhoCind}
 \bar\rho^{C}_{\mathcal{B}_k} \;=&\; \sum_{m=2}^k(-1)^{(k-m)}\sum_{\mathcal{C}_m\subseteq\mathcal{B}_k}
 \left(\bar\rho_{\mathcal{C}_m}\prod_{j\in \mathcal{B}_k\backslash \mathcal{C}_m}\bar\rho_j\right) \nonumber\\ & \qquad  -(-1)^k(k-1)\prod_{j\in  \mathcal{B}_k}\bar\rho_j.
\end{align}
Substituting equation \eqref{eq:rhoCind} into
\begin{equation}
\label{eq:rhoAnind}
\bar{\rho}_{\mathcal{A}_n}=\prod_{j\in\set{A}_n}\bar\rho_j+
\bar\rho_{\set{A}_n}^C+\sum_{k=2}^{n-1}
\sum_{\mathcal{C}_k\subseteq \mathcal{A}_n}\left(\prod_{j\in\mathcal{A}_n\backslash\set{C}_k}\bar{\rho}_{j}\right)
\bar{\rho}^{C}_{\mathcal{C}_k}
\end{equation}
then gives an expression of the form
\begin{equation}
\bar{\rho}_{\mathcal{A}_n}^C=\bar{\rho}_{\mathcal{A}_n}+\sum_{\ell=2}^{n-1}
\sum_{\set{F}_\ell\subset\set{A}_n}\xi_{\ell}\bar\rho_{\set{F}_\ell}
\prod_{i\in\set{A}_n\backslash\set{F}_\ell}\bar\rho_i+\xi_0\prod_{i\set{A}_n}\bar\rho_i.
\end{equation}

This is because terms in \eqref{eq:rhoAnind} contain one $\bar\rho^C_{\set{C}_k}$ multiplied by the single cell reduced density matrices for the rest of the cells, and terms in \eqref{eq:rhoCind} contain one reduced density matrix over a larger set multiplied by single cell reduced density matrices, and all subsets of the same size appear symmetrically in \eqref{eq:rhoAnind} and \eqref{eq:rhoCind}. Thus the final expression is a sum over terms which are the product of a single reduced density matrix over a set $\set{F}_\ell\subseteq\set{A}_n$ multiplied by  single cell reduced density matrices with a coefficient depending only on the size $\ell$ of the set $\set{F}_\ell$. Now we need to find $\xi_\ell$ and $\xi_0$.

To find $\xi_{\ell}$ we note that every $\set{B}_k\supseteq\set{F}_\ell$ ($\set{B}_k\subset\set{A}_n$) gives a contribution $-(-1)^{k-\ell}$ to $\xi_\ell$, so there are ${}^{n-\ell}C_{k-\ell}$ such  $\set{B}_k$'s for a given $k$; thus the coefficient is
\begin{align}
\xi_\ell=& \; -\sum_{k=\ell}^{n-1}(-1)^{k-\ell}\left({}^{n-\ell}C_{k-\ell}\right)= \;
-\sum_{p=0}^{n-\ell-1}(-1)^{p}\left({}^{n-\ell}C_{p}\right) \nonumber\\
=& \; -\sum_{p=0}^{n-\ell}(-1)^{p}\left({}^{n-\ell}C_{p}\right)+(-1)^{n-\ell}
\left({}^{n-\ell}C_{n-\ell}\right) \nonumber\\ =& \; -(1-1)^{n-\ell}+(-1)^{n-\ell}  \nonumber\\
=&\;\;\; (-1)^{n-\ell}
\end{align}
as required.

To find $\xi_0$ we note that there is a contribution $-1$ from the first term in equation \eqref{eq:rhoAnind} as well as a contribution $(-1)^k(k-1)$ from every $\set{B}_k$ with $n-1\geq k\geq 2$. There are ${}^nC_k$ different $\set{B}_k$s for each $k$, so that
\begin{align}
\xi_0\;=& \; -1+\sum_{k=2}^{n-1}(-1)^k(k-1)\left({}^nC_k\right)  \nonumber\\ \; =& \;\;\; (-1)^{n+1}(n-1)
\end{align}
as required; this completes the proof.


\subsection{Proof that any partial trace of $\bar{\rho}^C_{\set{A}_n}$ is zero}
 \label{sec:App-A2}


In the main text we took the result in eqtn. (\ref{eq:trrhoc0}) to be a defining property of the partial trace.  However, one can also derive the result explicitly from the expression \eqref{eq:rhoC}. We now show this.

Let us begin with (\ref{eq:rhoC}) of the main text, viz.,
\begin{align}
\tr{i}\bar\rho^{C}_{\mathcal{A}_n} \;=&\; \sum_{m=2}^n(-1)^{(n-m)}
\sum_{\mathcal{C}_m\subseteq\mathcal{A}_n}\tr{i}\left(\bar\rho_{\mathcal{C}_m}\prod_{j\in \mathcal{A}_n\backslash \mathcal{C}_m}\bar\rho_j\right) \nonumber\\ & \qquad   -(-1)^n(n-1)\prod_{j\in  \mathcal{A}_n\backslash{i}}\bar\rho_j.
\end{align}
with the notation as before.

We start by noting that
\begin{equation}
\tr{i}\left(\bar\rho_{\mathcal{C}_m}\prod_{j\in \mathcal{A}_n\backslash \mathcal{C}_m}\bar\rho_j\right)=\begin{cases}\displaystyle{\bar\rho_{\mathcal{C}_m}\prod_{j\in (\mathcal{A}_n\backslash i)\backslash \mathcal{C}_m}\bar\rho_j\quad i\not\in\set{C}_m}\\
\displaystyle{\bar\rho_{\mathcal{C}_m\backslash i}\prod_{j\in \mathcal{A}_n\backslash \mathcal{C}_m}\bar\rho_j\quad i\in\set{C}_m}
\end{cases}
\end{equation}

\begin{widetext}
It then follows that we can write
\begin{align}
\sum_{m=2}^n(-1)^{(n-m)}\sum_{\mathcal{C}_m\subseteq\mathcal{A}_n}\tr{i}
\left(\bar\rho_{\mathcal{C}_m}\prod_{j\in \mathcal{A}_n\backslash \mathcal{C}_m}\bar\rho_j\right) \; =& \; \sum_{\ell\in\set{A}_n\backslash i}\tr{i}\bar\rho_{i\ell}(-1)^{n-2}\prod_{j\in \mathcal{A}_n\backslash \{i,\ell\}}\bar\rho_j
\nonumber\\ & \qquad \;+\; \sum_{m=2}^{n-2}
\sum_{\set{C}_{m}\subseteq(\set{A}_n\backslash i)}(-1)^m\tr{i}\left(\bar\rho_{\set{C}_m}\bar\rho_i-\bar\rho_{\set{C}_m\cup\{i\}}\right)
\prod_{j\in(\set{A}_n\backslash i)\backslash \set{C}_m}\bar\rho_j
\nonumber\\
=& \; \sum_{\ell\in\set{A}_n\backslash i}(-1)^{n-2}\prod_{j\in \mathcal{A}_n\backslash \{i\}}\bar\rho_j  \nonumber\\   =& \;\;\;  (n-1)(-1)^{n-2}\prod_{j\in \mathcal{A}_n\backslash \{i\}}\bar\rho_j
\end{align}
so that
\begin{align}
\tr{i}\bar\rho^{C}_{\mathcal{A}_n} \;=&\;\;\;(n-1)(-1)^{n-2}
\prod_{j\in \mathcal{A}_n\backslash \{i\}}\bar\rho_j-(-1)^n(n-1)\prod_{j\in  \mathcal{A}_n\backslash{i}}\bar\rho_j \nonumber \\      =&\;\; \;0
\end{align}
which is the result we wanted.


\section{Derivation of Equations of Motion hierarchies}
 \label{sec:App-B}


In the main text we simply quoted the results for the equations of motion, for both a general multipartite system, and also for an $N$-qubit system. Here we give the derivations of these results.

\subsection{Equation of Motion for $N$-partite system}
 \label{sec:App-B1}

Write begin by writing the Hamiltonian as a "free" single-system part, plus a pairwise interaction term, viz.,
\begin{equation}
H=H^0+H^I=\sum_j\left(H^0_j+\frac{1}{2}\sum_{i\neq j}H_{ij}^I\right)
\end{equation}

The equation of motion is then
\begin{align}
i\partial_t\rho_\set{S} \;&=\;[H,\rho_\set{S}] \nonumber \\
&=\;\sum_{\mathcal{A}\subseteq \mathcal{S}}\Biggl[H\,,\,\left(\prod_{j\not\in\mathcal{A}}\bar{\rho}_{j}\right)
\bar{\rho}^{C}_{\mathcal{A}} \Biggr] \nonumber \\
&=\;\sum_{j\in\set{S}}\left\{\left[H_j^0+
\sum_{j\neq i\in\set{S}}\frac{1}{2}H_{ij}^0\,,\,\left(\prod_{j\not\in\mathcal{A}}\bar{\rho}_{j}\right)
\bar{\rho}^{C}_{\mathcal{A}}\right]\right\}
\end{align}
for the part of the above containing the non-interacting part of the Hamiltonian each $j$ is either in $\set{A}$ or not $\set{A}$, for the interacting part there are three possible situations (see figure \ref{fig:vennint}): Both $i,j\in\set{A}$,  only one of $i$ or $j$ in $\set{A}$, and both $i,j\notin\set{A}$. We can split the sums up accordingly; one has
\begin{align}
\sum_{\set{A}}\left[H,\,\left(\prod_{j\notin\mathcal{A}}\bar{\rho}_{j}\right)
\bar{\rho}^{C}_{\mathcal{A}}\right]=&\sum_{\set{A}\subseteq{S}}\Biggl\{\sum_{j\in\set{A}}
\left[H_{j}^0,\bar{\rho}^{C}_{\set{A}}\right] \prod_{i\notin\set{A}}\bar{\rho}_i+\bar{\rho}^{C}_{\set{A}}
\sum_{j\notin\set{A}}\left[H^0_{j},\bar{\rho}_{j}\right]
\prod_{i\notin\set{A}\cup\{j\}}\bar\rho_{i}   \nonumber\\
&+\sum_{j\in\set{A}}\sum_{i\in\set{A}\backslash\{j\}}
\left[\tfrac{1}{2}H_{ij}^I,\,\bar\rho_{\set{A}}^C\right] \prod_{k\notin\set{A}}\bar{\rho}_k+\sum_{j\in\set{A}}
\sum_{i\notin\set{A}}\left[\tfrac{1}{2}H_{ij}^I,\,\bar\rho_{\set{A}}^C\bar{\rho}_i\right] \prod_{k\notin\set{A}\cup\{i\}}\bar{\rho}_k\nonumber\\
&  +\bar\rho^C_\set{A}\sum_{j\notin\set{A}}\sum_{i\notin\set{A}\cup\{j\}}
\left[\tfrac{1}{2}H_{ij}^I,\,\bar\rho_j\bar\rho_i\right]\prod_{k\notin
\set{A}\cup\{i,j\}}\bar{\rho}_k\Biggr\}.\label{eq:apcomutator}
\end{align}


\begin{figure}\label{fig:vennint}
  \caption{The different classes of interaction involving $\set{A}$. In (i) we have interactions entirely between cells inside $\set{A}$; in (ii) we have interactions between clls inside $\set{A}$ and cells outside; and in (ii) the interactions are entirely between cells outside $\set{A}$. The interactions
  are denoted by the wavy line.}
  \centering
    \includegraphics[width=0.5\textwidth]{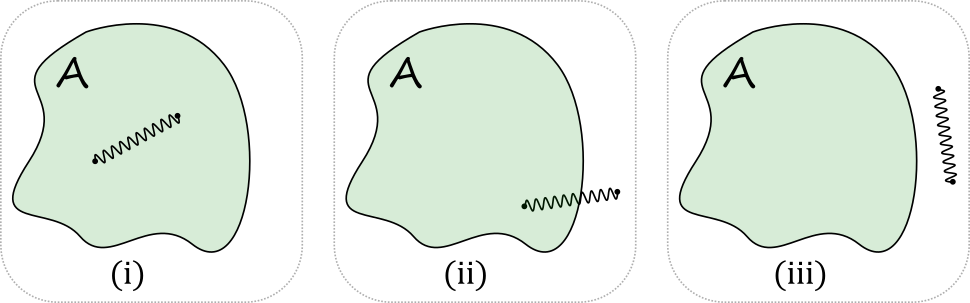}
\end{figure}


Now we trace out a set $\set{C}$ of cells. This gives
\begin{align}
\tr{\set{C}}\left[H,\,\sum_{\set{A}}\bar{\rho}_{\set{A}}^C\prod_{i\notin \set{A}}\bar{\rho}_i\right]=&\sum_{\set{A}}\tr{\set{C}}\sum_{j\in\set{A}}
\left[H_{j}^0,\bar{\rho}^{C}_{\set{A}}\right] \prod_{i\notin\set{A}}\bar{\rho}_i+\sum_{\set{A}}
\tr{\set{C}}\bar{\rho}^{C}_{\set{A}}\sum_{j\notin\set{A}}
\left[H^0_{j},\bar{\rho}_{j}\right]\prod_{i\notin\set{A}\cup\{j\}}\bar\rho_{i}
 \nonumber\\
&+\sum_{\set{A}}\sum_{j\in\set{A}}\sum_{i\in\set{A}\backslash\{j\}}
\tr{\set{C}}\left[\tfrac{1}{2}H_{ij}^I,\,\bar\rho_{\set{A}}^C\right] \prod_{k\notin\set{A}}\bar{\rho}_k +\sum_{\set{A}}\sum_{j\in\set{A}}\tr{\set{C}}\sum_{i\notin\set{A}}
\left[\tfrac{1}{2}H_{ij}^I,\,\bar\rho_{\set{A}}^C\bar{\rho}_i\right] \prod_{k\notin\set{A}\cup\{i\}}\bar{\rho}_k\nonumber\\
& +\sum_{\set{A}}\tr{\set{C}}\bar\rho^C_\set{A}\sum_{j\notin\set{A}}
\sum_{i\notin\set{A}\cup\{j\}}\left[\tfrac{1}{2}H_{ij}^I,\,\bar\rho_j
\bar\rho_i\right]\prod_{k\notin\set{A}\cup\{i,j\}}\bar{\rho}_k.
 \label{eq:trcomrho}
\end{align}

Let us simplify each term in the above equation separately:
\begin{enumerate}
\item First consider the terms involving $H^0_j$:
\begin{enumerate}
\item Consider the first sum in eqtn. \eqref{eq:trcomrho}, viz.,
\begin{equation}
\sum_{\set{A}}\tr{\set{C}}\sum_{j\in\set{A}}\left[H_{j}^0,\bar{\rho}^{C}_{\set{A}}\right] \prod_{i\notin\set{A}}\bar{\rho}_i
\end{equation}
\begin{itemize}
\item The terms are trivially zero when the overlap $\set{C} \cap \set{A}\neq\varnothing$ or $\{j\}$.
\item If the overlap contains exactly one cell $\set{C} \cap \set{A}=\{j\}$, then we have the following identity
\begin{equation*}
\tr{j}\left[H_{j}^0,\bar{\rho}^{C}_{\set{A}}\right]=(H_j^0)_{mn}
(\bar{\rho}^{C}_{\set{A}})_{mn}^{MM}-(H_j^0)_{nm}(\bar{\rho}^{C}_{\set{A}})_{mn}^{MM}=0.
\end{equation*}
\end{itemize}
where $M$ is an index on the Hilbert space of states on the set of cells $\set{A}\backslash\{j\}$ and $m,n$ are indices on the Hilbert space at $j$, and repeated indices are summed so $(\bar{\rho}^{C}_{\set{A}})_{mn}^{MM}=\sum_M\langle m M|(\bar{\rho}^{C}_{\set{A}})_{mn}^{MM}|nM\rangle$.

We see therefore that only terms with no overlap $\set{C} \cap \set{A}=\varnothing$ contribute to the first sum in eqtn. \eqref{eq:trcomrho}:
\begin{equation}\label{eqaprhosum1}
 \sum_{\set{A}\subseteq\set{S}}\tr{\set{C}}\sum_{j\in\set{A}}
 \left[H_{j}^0,\bar{\rho}^{C}_{\set{A}}\right] \prod_{i\notin\set{A}}\bar{\rho}_i=\sum_{\set{A}\subseteq( \set{S}\backslash\set{C})}\sum_{j\in\set{A}}\left[H_{j}^0,\bar{\rho}^{C}_{\set{A}}\right]
 \tr{\set{C}} \prod_{i\notin\set{A}}\bar{\rho}_i=\sum_{\set{A}\subseteq \set{S}\backslash\set{C}}\sum_{j\in\set{A}}\left[H_{j}^0,\bar{\rho}^{C}_{\set{A}}\right] \prod_{i\in\set{S}\backslash(\set{A}\cup\set{C})}\bar{\rho}_i
 \end{equation}
\item Consider now the second sum in \eqref{eq:trcomrho}, viz.,
$$\tr{\set{C}}\bar{\rho}^{C}_{\set{A}}\sum_{j\notin\set{A}}^N\left[H^0_{j},
\bar{\rho}_{j}\right]\prod_{i\notin\set{A}\cup\{j\}}\bar\rho_{i}$$
The terms are zero when $\set{A}\cap\set{C}\neq\varnothing$ and when $j\in\set{C}$, so that
\begin{align}
\sum_{\set{A}\subseteq\set{S}}\tr{\set{C}}\bar{\rho}^{C}_{\set{A}}
\sum_{i\notin\set{A}}\left[H^0_{j},\bar{\rho}_{j}\right]
\prod_{j\notin\set{A}\cup\{j\}}\bar\rho_{i}&=
\sum_{\set{A}\subseteq(\set{S}\backslash\set{C})}\bar{\rho}^{C}_{\set{A}_n}
\sum_{j\notin(\set{A}\cup\set{C})}\left[H^0_{j},\bar{\rho}_{j}\right]
\tr{\set{C}}\prod_{i\notin\set{A}\cup\{j\}}\bar\rho_{i}\nonumber\\&=
\sum_{\set{A}\subseteq(\set{S}\backslash\set{C})}\sum_{j\notin(\set{A}\cup\set{C})}
\bar{\rho}^{C}_{\set{A}}\left[H^0_{j},\bar{\rho}_{j}\right]
\prod_{i\in\set{S}\backslash(\set{A}\cup\set{C}\cup\{j\})}
\bar\rho_{i}
 \nonumber\\
&=\sum_{\set{A}\subseteq(\set{S}\backslash\set{C})}\sum_{j\in\set{A}}
\bar{\rho}^{C}_{\set{A}\backslash\{j\}}\left[H^0_{j},\bar{\rho}_{j}\right]
\prod_{i\in\set{S}\backslash(\set{A}\cup\set{C})}\bar\rho_{i}
 \label{eqaprhosum2}
\end{align}
The last line here requires a bit of thought; it reflects the fact that summing over all possible $\set{A}\subseteq(\set{S}\backslash\set{C})$, then over $j\in(\set{S}\backslash(\set{C}\cup\set{A})$, is equivalent to summing over all possible $\set{A}\subseteq(\set{S}\backslash\set{C})$ and all possible $j$ in $\set{A}$.
\end{enumerate}

\vspace{3mm}

\item Now consider the terms involving the interaction Hamiltonian $H_{ij}^I$.
\begin{enumerate}
\item  Consider first the third sum in equation \eqref{eq:trcomrho}, viz.,
\begin{equation}
\sum_{\set{A}}\sum_{j\in\set{A}}\sum_{i\in\set{A}\backslash\{j\}}
\tr{\set{C}}\left[\tfrac{1}{2}H_{ij}^I,\,\bar\rho_{\set{A}}^C\right] \prod_{k\notin\set{A}}\bar{\rho}_k\label{apsum3}
\end{equation}
which contains all the terms where cells inside $\set{A}$ are interacting with each other, ie., case (i) in figure \ref{fig:vennint}.
\begin{itemize}
\item If the intersection $\set{C}\cap\set{A}$ contains cells other than $i$ or $j$, then $$\tr{\set{C}}\left[\tfrac{1}{2}H_{ij}^I,\,\bar\rho_{\set{A}}^C\right] \prod_{k\notin\set{A}}\bar{\rho}_k=0.$$
\item If the intersection $\set{C}\cap\set{A}=\{i,j\}$  then $$\tr{\set{C}}\left[\tfrac{1}{2}H_{ij}^I,\,\bar\rho_{\set{A}}^C\right] \prod_{k\notin\set{A}}\bar{\rho}_k=0.$$
so that there are only nonzero terms in the sum when the intersection $\set{A}\cap\set{C}$ contains exactly one or zero elements.
\item  If the intersection is one of $\set{C}\cap\set{A}=\{i\}$ or $\{j\}$ then $$\tr{\set{C}}\left[\tfrac{1}{2}H_{ij}^I,\,\bar\rho_{\set{A}}^C\right] \prod_{k\notin\set{A}}\bar{\rho}_k=\tr{i\text{or}j}
    \left(\left[\tfrac{1}{2}H_{ij}^I,\,\bar\rho_{\set{A}}^C\right]\right) \prod_{k\notin(\set{A}\cup\set{C})}\bar{\rho}_k$$  which is not necessarily zero.
\item When both $i$, $j$ are in $\set{A}\backslash\set{C}$ then
$$\tr{\set{C}}\left[\tfrac{1}{2}H_{ij}^I,\,\bar\rho_{\set{A}}^C\right] \prod_{k\notin\set{A}}\bar{\rho}_k=\left[\tfrac{1}{2}H_{ij}^I,\,\bar\rho_{\set{A}}^C\right] \prod_{k\notin(\set{A}\cup\set{C})}\bar{\rho}_k.$$
\end{itemize}
Thus there only two kinds of term in the sum \eqref{apsum3} that matter. The first are those where both $i,j\not\in\set{C}$ and $\set{C}\cap\set{A}$. The second are those where only one of $i$ or $j$ are in $\set{C}$ (say $i\in\set{C}$) and $\set{A}\cap\set{C}=\{i\}$. Thus
\begin{align}
\sum_{\set{A}}\sum_{j\in\set{A}}\sum_{i\in\set{A}\backslash\{j\}}
\left[\tfrac{1}{2}H_{ij}^I,\,\bar\rho_{\set{A}}^C\right] \tr{\set{C}}\prod_{k\notin\set{A}}\bar{\rho}_k=
&\sum_{\set{A}\subseteq(\set{S}\backslash\set{C})}
\sum_{j\in\set{A}}\sum_{i\in\set{A}\backslash\{j\}}\tr{\set{C}}
\left[\tfrac{1}{2}H_{ij}^I,\,\bar\rho_{\set{A}}^C\right] \prod_{k\notin\set{A}}\bar{\rho}_k\nonumber\\&+
\sum_{\substack{\set{A}\subseteq\set{S}\\\set{A}\cap\set{C}=
\{j\}}}\sum_{j\in\set{A}}\tr{\set{C}}\left[\tfrac{1}{2}H_{ij}^I,\,\bar\rho_{\set{A}}^C\right] \prod_{k\notin\set{A}}\bar{\rho}_k
 \nonumber\\
=&\sum_{\set{A}\subseteq(\set{S}\backslash\set{C})}\sum_{j\in\set{A}}
\sum_{i\in\set{A}\backslash\{j\}}\left[\tfrac{1}{2}H_{ij}^I,\,\bar\rho_{\set{A}}^C\right] \prod_{k\notin\set{A}\cup\set{C}}\bar{\rho}_k
 \nonumber\\
 &+\sum_{\set{A}\subseteq(\set{S}\backslash\set{C})}\sum_{i\in\set{C}}
 \sum_{j\in\set{A}}\left[H_{ij}^I,\,\bar\rho_{\set{A}\cup\{i\}}^C\right] \prod_{k\notin\set{A}\cup\set{C}}\bar{\rho}_k
  \label{eqaprhosum3}
\end{align}
\item The fourth sum in eqtn. \eqref{eq:trcomrho}, viz.,
\begin{equation}
\sum_{\set{A}}\sum_{j\in\set{A}}\tr{\set{C}}\sum_{i\notin\set{A}}
\left[\tfrac{1}{2}H_{ij}^I,\,\bar\rho_{\set{A}}^C\bar{\rho}_i\right] \prod_{k\notin\set{A}\cup\{i\}}\bar{\rho}_k
 \label{apsum4}
\end{equation}
is a sum over terms involving interactions between $j$ in $\set{A}$ and $i$ not in $\set{A}$ (ie., terms like (ii) in figure \ref{fig:vennint}).
\begin{itemize}
\item When neither $i$ nor $j$ are in $\set{C}$, then
$$\tr{\set{C}}\left[\tfrac{1}{2}H_{ij}^I,\,\bar\rho_{\set{A}}^C\bar{\rho}_i\right] \prod_{k\notin\set{A}\cup\{i\}}\bar{\rho}_k=
\left[\tfrac{1}{2}H_{ij}^I,\,\bar\rho_{\set{A}}^C\bar{\rho}_i\right] \prod_{k\notin(\set{A}\cup\set{C}\cup\{i\})}\bar{\rho}_k.$$
\item When there is an overlap $\set{A}\cup\set{C}$ which contains an element other than $j$, then
$$\tr{\set{C}}\left[\tfrac{1}{2}H_{ij}^I,\,\bar\rho_{\set{A}}^C\bar{\rho}_i\right] \prod_{k\notin\set{A}\cup\{i\}}\bar{\rho}_k=0.$$
\item When $i$ is in $\set{C}$ and $\set{A}\cap\set{C}=\varnothing$, then
$$\tr{\set{C}}\left[\tfrac{1}{2}H_{ij}^I,\,\bar\rho_{\set{A}}^C\bar{\rho}_i\right] \prod_{k\notin\set{A}\cup\{i\}}\bar{\rho}_k=\left[\tfrac{1}{2}\tr{i}
\left(H_{ij}^I\bar{\rho}_i\right),\,\bar\rho_{\set{A}}^C\right] \prod_{k\notin(\set{A}\cup\set{C}\cup\{i\})}\bar{\rho}_k.$$
\item When $j$ is in $\set{C}$ but $i$ is not, then $$\tr{\set{C}}\left[\tfrac{1}{2}H_{ij}^I,\,\bar\rho_{\set{A}}^C\bar{\rho}_i\right] \prod_{k\notin\set{A}\cup\{i\}}\bar{\rho}_k=
    \tr{j}\left[\tfrac{1}{2}H_{ij}^I,\,\bar\rho_{\set{A}}^C\bar{\rho}_i\right] \prod_{k\notin(\set{A}\cup\set{C}\cup\{i\})}\bar{\rho}_k.$$
\item If $i$ and $j$ are in $\set{C}$, then
$$\tr{\set{C}}\left[\tfrac{1}{2}H_{ij}^I,\,\bar\rho_{\set{A}}^C\bar{\rho}_i\right] \prod_{k\notin\set{A}\cup\{i\}}\bar{\rho}_k=0.$$
\end{itemize}
Thus the sum \eqref{apsum4} is
\begin{align}
\sum_{\set{A}}\sum_{j\in\set{A}}\tr{\set{C}}\sum_{i\notin\set{A}}
\left[\tfrac{1}{2}H_{ij}^I,\,\bar\rho_{\set{A}}^C\bar{\rho}_i\right] \prod_{k\notin\set{A}\cup\{i\}}\bar{\rho}_k=&\sum_{\set{A}
\subset(\set{S}\backslash\set{C})}\sum_{j\in\set{A}}
\sum_{i\notin(\set{A}\cup\set{C})}\left[\tfrac{1}{2}H_{ij}^I,\,
\bar\rho_{\set{A}}^C\bar{\rho}_i\right] \prod_{k\notin(\set{A}\cup\set{C}\cup\{i\})}\bar{\rho}_k
 \nonumber\\
&+\sum_{\set{A}\subset(\set{S}\backslash\set{C})}\sum_{j\in\set{A}}
\sum_{i\in\set{C}}\left[\tfrac{1}{2}\tr{i}\left(H_{ij}^I\bar{\rho}_i\right),
\,\bar\rho_{\set{A}}^C\right] \prod_{k\notin(\set{A}\cup\set{C})}\bar{\rho}_k
 \nonumber\\
&+\sum_{\substack{\set{A}\subseteq\set{S}\\\set{A}\cap\set{C}=
\{j\}}}\sum_{i\notin(\set{A}\cup\set{C})}\tr{j}\left[\tfrac{1}{2}H_{ij}^I,
\,\bar\rho_{\set{A}}^C\bar{\rho}_i\right] \prod_{k\notin(\set{A}\cup\set{C}\cup\{i\})}\bar{\rho}_k\\
=&\sum_{\set{A}\subset(\set{S}\backslash\set{C})}\sum_{j\in\set{A}}
\sum_{i\notin(\set{A}\cup\set{C})}\left[\tfrac{1}{2}H_{ij}^I,\,
\bar\rho_{\set{A}}^C\bar{\rho}_i\right] \prod_{k\notin(\set{A}\cup\set{C}\cup\{i\})}\bar{\rho}_k
 \nonumber\\
&+\sum_{\set{A}\subset(\set{S}\backslash\set{C})}\sum_{j\in\set{A}}
\sum_{i\in\set{C}}\left[\tfrac{1}{2}\tr{i}\left(H_{ij}^I\bar{\rho}_i\right),\,
\bar\rho_{\set{A}}^C\right] \prod_{k\notin(\set{A}\cup\set{C})}\bar{\rho}_k
 \nonumber\\
&+\sum_{\set{A}\subseteq(\set{S}\backslash\set{C})}\sum_{j\in\set{C}}\tr{j}
\left[\tfrac{1}{2}H_{ij}^I,\,\bar\rho_{\set{A}\cup\{j\}}^C\bar{\rho}_i\right] \prod_{k\notin(\set{A}\cup\set{C}\cup\{i\})}\bar{\rho}_k
 \label{eqaprhosum4}
\end{align}

\item The fifth sum in eqtn. \eqref{eq:trcomrho}, viz.,
\begin{equation}
\sum_{\set{A}}\tr{\set{C}}\bar\rho^C_\set{A}\sum_{j\notin\set{A}}
\sum_{i\notin\set{A}\cup\{j\}}\left[\tfrac{1}{2}H_{ij}^I,\,\bar\rho_j\bar\rho_i\right]
\prod_{k\notin\set{A}\cup\{i,j\}}\bar{\rho}_k
 \label{apsum5}
\end{equation}
is a sum over the interactions shown in figure \ref{fig:vennint}
(iii), where both $i$ and $j$ are not in $\set{C}$. Then
\begin{itemize}
\item When $i,j\in\set{C}$, we have
$$\tr{\set{C}}\bar\rho^C_\set{A}\left[\tfrac{1}{2}H_{ij}^I,\,\bar\rho_j\bar\rho_i\right]
\prod_{k\notin\set{A}\cup\{i,j\}}\bar{\rho}_k=0.$$
\item When $\set{A}\cap\set{C}\neq\varnothing$, we have
$$\tr{\set{C}}\bar\rho^C_\set{A}\left[\tfrac{1}{2}H_{ij}^I,\,\bar\rho_j\bar\rho_i\right]
\prod_{k\notin\set{A}\cup\{i,j\}}\bar{\rho}_k=0.$$
\item When one of $i$ and $j$ (say $i$) is in $\set{C}$ and the other is not, then
(and $\set{A}\cap\set{C}=\varnothing$)
$$\tr{\set{C}}\bar\rho^C_\set{A}\left[\tfrac{1}{2}H_{ij}^I,\,
\bar\rho_j\bar\rho_i\right]\prod_{k\notin\set{A}\cup\{i,j\}}
\bar{\rho}_k=\bar\rho^C_\set{A}\left[\tfrac{1}{2}\tr{i}(H_{ij}^I\bar\rho_i),\,
\bar\rho_j\right]\prod_{k\notin(\set{A}\cup\{j\}\cup\set{C})}\bar{\rho}_k.$$
\item When nether $i$ nor $j$ are in $\set{C}$ and $\set{A}\cap\set{C}=\varnothing$, we have
$$\tr{\set{C}}\bar\rho^C_\set{A}\left[\tfrac{1}{2}H_{ij}^I,\,\bar\rho_j\bar\rho_i\right]
\prod_{k\notin\set{A}\cup\{i,j\}}\bar{\rho}_k=\bar\rho^C_\set{A}
\left[\tfrac{1}{2}H_{ij}^I,\,\bar\rho_j\bar\rho_i\right]
\prod_{k\notin\set{A}\cup\{i,j\}\cup\set{C}}\bar{\rho}_k.$$
\end{itemize}
Thus the sum \eqref{apsum5} is given by
\begin{align}
\sum_{\set{A}}\tr{\set{C}}\bar\rho^C_\set{A}\sum_{j\notin\set{A}}
\sum_{i\notin\set{A}\cup\{j\}}\left[\tfrac{1}{2}H_{ij}^I,\,
\bar\rho_j\bar\rho_i\right]\prod_{k\notin\set{A}\cup\{i,j\}}\bar{\rho}_k=
&\sum_{\set{A}\subseteq\set{C}}\sum_{i\in\set{C}}\sum_{j\notin\set{A}\cup\set{C}}
\bar\rho^C_\set{A}\left[\tfrac{1}{2}\tr{i}(H_{ij}^I\bar\rho_i),\,\bar\rho_j\right]
\prod_{k\notin(\set{A}\cup\{j\}\cup\set{C})}\bar{\rho}_k
 \nonumber\\
&+\sum_{\set{A}\subseteq\set{C}}\sum_{i\in\set{C}}\sum_{j\in\set{A}\backslash\{i\}}
\bar\rho^C_\set{A}\left[\tfrac{1}{2}H_{ij}^I,\,\bar\rho_j\bar\rho_i\right]
\prod_{k\notin\set{A}\cup\{i,j\}\cup\set{C}}\bar{\rho}_k.
 \label{eqaprhosum5}
\end{align}
\end{enumerate}
\end{enumerate}
Thus finally, inserting equations \eqref{eqaprhosum1},\eqref{eqaprhosum2},\eqref{eqaprhosum3},\eqref{eqaprhosum4}, and \eqref{eqaprhosum5} into \eqref{eq:trcomrho}, we have
\begin{align} \tr{\set{C}}\left[H,\,\rho_\set{S}\right]=&
\sum_{\set{A}\subseteq\set{S}\backslash\set{C}}\Biggl\{\sum_{j\in\set{A}}
\left[H_{j}^0,\bar{\rho}^C_{\set{A}}\right]
\prod_{i\in \set{S}\backslash (\set{A}\cup \set{C})}\bar{\rho}_i+\bar{\rho}^C_{\set{A}}\sum_{i\in \set{S}\backslash (\set{A}\cup \set{C})}\left[H^0_{i},\bar{\rho}_{i}\right]\prod_{j\in \set{S}\backslash(\set{A}\cup\set{C}\cup\{i\})}\bar\rho_{j}
 \nonumber \\
& +\sum_{j\in\set{A}}\sum_{i\in\set{A}\backslash\{j\}}
\left[\tfrac{1}{2}H_{ij}^I,\bar{\rho}^C_{\set{A}_n}\right] \prod_{k\in \set{S}\backslash(\set{A}\cup\set{C})}\bar{\rho}_k+ \sum_{i\in\set{C}}\sum_{j\in\set{A}}\text{tr}_{i} \left[H_{ij}^I,\bar{\rho}^C_{\set{A}\cup\{i\}}\right] \prod_{k\in \set{S}\backslash(\set{A}\cup \set{C})}\bar{\rho}_k
 \nonumber\\
&+\bar{\rho}^C_{\set{A}}\sum_{k\in \set{S}\backslash(\set{A}\cup \set{C})}\sum_{l\in \set{S}\backslash(\set{A}\cup \set{C}\cup\{k\})}\left[H^I_{kl},\bar{\rho}_k\right]\prod_{j\in \set{S}\backslash(\set{A}\cup \set{C}\cup\{k,l\})}\bar\rho_j
 \nonumber\\
& +\bar{\rho}^C_{\set{A}}\sum_{k\in\set{S}\backslash(\set{A}\cup \set{C})}\sum_{l\in\set{C}}\left[\tr{l} (H^I_{kl}\bar{\rho}_l),\bar{\rho}_k\right]\prod_{j\in \set{S}\backslash(\set{A}\cup \set{C}\cup\{k\})}\bar\rho_j
 \nonumber \\
& +\sum_{j\in\set{A}}\sum_{k\in \set{S}\backslash (\set{A}\cup \set{C})}\left[H_{jk}^I,\bar{\rho}^{C}_{\set{A}}\bar\rho_k\right] \prod_{i\in\set{S}\backslash (\set{A}\cup \set{C}\cup\{k\})}\bar{\rho}_i
 \nonumber\\
&+ \sum_{l\in\set{C}}\sum_{k\in \set{S}\backslash (\set{A}\cup \set{C})}\text{tr}_{l}\left[H_{lk}^I,\bar{\rho}^{C}_{\set{A}\cup\{l\}}\bar\rho_k\right] \prod_{j\in \set{S}\backslash (\set{A}\cup \set{C}\cup\{k\})}\bar{\rho}_j
 \nonumber\\
& + \sum_{l\in\set{C}}\sum_{k\in \set{A}}\text{tr}_{l}
\left[ H_{jl}^I\bar\rho_l,\bar{\rho}^{C}_{\set{A}}\right] \prod_{j\in \set{S}\backslash (\set{A}\cup \set{C})}\bar{\rho}_j\Biggr\}.
\end{align}
Comparing this with equation \eqref{eq:apcomutator}, we see that all of those terms above which do not contain an explicit trace can be collected to give $\left[H_{\set{S}\backslash\set{C}},\,\bar{\rho}_{\set{S}\backslash\set{C}}\right]$, with
\begin{equation}
H_{\set{S}\backslash\set{C}}=\sum_{j\in\set{S}\backslash\set{C}}
\left(H^0_j+\frac{1}{2}\sum_{j\in\set{S}\backslash(\set{C}\cup\{j\})}H_{ij}^I\right)
\end{equation}
 so that
 \begin{equation}
 i\partial_t\bar\rho_{\set{S}\backslash\set{C}}=
 \left[H_{\set{S}\backslash\set{C}},\,\bar{\rho}_{\set{S}\backslash\set{C}}\right]+
 \textrm{TT}
 \end{equation}
The extra ``trace term" $TT$ is
\begin{align}
\textrm{TT}=&\; i\sum_{l\in\set{C}}\text{tr}_{l}\left\{\sum_{\set{A}\subseteq (\set{S}\backslash\set{C})}\sum_{i\in\set{A}}\left( \left[H^I_{il},\bar\rho_{\set{A}}^C\right]\bar\rho_l+\left[H^I_{il}\, ,\, \rho^{C}_{\set{A}\cup\{l\}}\right]+\left[H^I_{il}\, ,\, \bar\rho_i\right] \rho^{C}_{(\set{A}\backslash \{i\})\cup\{l\}}+\left[H^I_{il}\, ,\, \bar\rho_i\right] \rho^{C}_{(\set{A}\backslash \{i\})}\bar\rho_l\right)\prod_{j\in (\set{S}\backslash\set{C})\backslash\set{A}} \bar\rho_j\right\}\\
 =&\; i\sum_{l\in\set{C}}\text{tr}_{l}
 \left[\sum_{{i\in{\set{S}\backslash\set{C}}}}H_{il}^I,\,\sum_{\set{A}\subseteq (\set{S}\backslash\set{C})\cup\{l\}}\rho^C_{\tilde{S}}\prod_{j\in ((\set{S}\backslash\set{C})\cup\{l\})\backslash\set{A}}\bar\rho_j\right] \;\;=\;\;
 \sum_{l\in\set{C}}\left[\sum_{i\in{\set{S}\backslash\set{C}}}H_{il}^I,\,
 \bar{\rho}_{(\set{S}\backslash\set{C})\cup\{l\}}\right]
\end{align}

Thus, finally, we have the result
\begin{equation}
i\partial_t\bar\rho_{\set{S}\backslash\set{C}}=
\left[H_{\set{S}\backslash\set{C}},\,\bar{\rho}_{\set{S}\backslash\set{C}}\right]+
i\sum_{l\in\set{C}}\left[\sum_{i\in{\set{S}\backslash\set{C}}}H_{il}^I,\,
\bar{\rho}_{(\set{S}\backslash\set{C})\cup\{l\}}\right].
\end{equation}

If we now relabel the set $(\set{S}\backslash{C})\to\set{A}$, we get the result (\ref{eq:rhoA}) in the text.
\end{widetext}

\subsection{Equation of Motion for $N$-qubit system}
 \label{sec:App-B2}

We now want to derive the equations of motion (\ref{eq:EOMgensp12}) for $N$ qubits. The Hamiltonian is
\begin{equation}
H \;=\;\sum_i\tfrac{1}{2}\vc{h}_i\cdot\boldsymbol{\sigma}_i+\sum_{i=1}^N\sum_{j<i}\tfrac{1}{2}V_{ij}^{\mu\nu}\sigma_i^\mu\sigma_j^\nu
\end{equation}
which we write as $H = H_0+H_V$. 

We wish to calculate
\begin{equation}
\frac{\ud}{\ud t}\Bigl\langle\prod_{j\in\set{C}}\sigma_j^{\mu_j}\Bigr\rangle=-i\biggl\langle\Bigl[H,\,\prod_{j\in\set{C}}\sigma_j^{\mu_j}\Bigr]\biggr\rangle.\label{eq:HEOMcor}
\end{equation}
We thus need the commutators
\begin{eqnarray}
\left[H_0,\prod_{j\in\set{C}}\sigma_j^{\mu_j}\right] \;&=&\;\sum_{i\in\set{C}}\frac{1}{2}h_i^\lambda[\sigma_i^{\lambda},\sigma_i^{\mu_i}]\prod_{j\in\set{C}\backslash\{i\}}\sigma_j^{\mu_j}    \nonumber \\ \;&=&\;
i\sum_{i\in\set{C}}\varepsilon^{\mu_i\lambda\nu_i}h_i^\lambda\sigma_i^{\nu_i}\prod_{j\in\set{C}\backslash\{i\}}\sigma_j^{\mu_j} \qquad
\end{eqnarray}
and
\begin{equation}
\left[H_V,\prod_{j\in\set{C}}\sigma_j^{\mu_j}\right]=\sum_{i=1}^N\sum_{k<i}\tfrac{1}{2}V_{ik}^{\alpha\beta}\left[\sigma_i^\alpha\sigma_k^\beta,\,\prod_{j\in\set{C}}\sigma_j^{\mu_j}\right] \;\;\; \qquad
\end{equation}

The commutator on the right of the previous expression is non zero when either one of $i,k$ or both $i$ and $k$ are in $\set{C}$. Consider the case when $i$ is in $\set{C}$ but $k$ is not; then we have
\begin{equation}
\left[\sigma_i^\alpha\sigma_k^\beta,\,\prod_{j\in\set{C}}\sigma_j^{\mu_j}\right]\;=\;2i\varepsilon^{\mu_i\alpha\nu_i}\sigma_i^{\nu_i}\sigma_k^\beta\prod_{j\in\set{C}\backslash{i}}\sigma_j^{\mu_j}\label{eq:com2}
\end{equation}
On the other hand if both $i$ and $k$ are in $\set{C}$, then we have \begin{widetext}
\begin{eqnarray}
 \label{eq:com1}
\left[\sigma_i^\alpha\sigma_k^\beta,\,\prod_{j\in\set{C}}\sigma_j^{\mu_j}\right] \;&=& \; \left[\sigma_i^\alpha\sigma_k^\beta,\sigma_i^{\mu_i}\sigma_k^{\mu_k}\right]\prod_{j\in\set{C}\backslash\{i,k\}}\sigma_j^{\mu_j} \nonumber \\
 \;&=&\; 2i\left(\varepsilon^{\mu_i\alpha\nu_i}\delta^{\mu_k\beta}\sigma_j^{\nu_j}+\varepsilon^{\mu_k\beta\nu_k}\sigma_k^{\nu_k}\delta^{\mu_j\alpha}\right)\prod_{j\in\set{C}\backslash\{i,k\}}\sigma_j^{\mu_j}
\end{eqnarray}
putting equations \eqref{eq:com2} and \eqref{eq:com2} into the equation of motion for the correlator \eqref{eq:HEOMcor} one gets the hierarchy of equations of motion,
\begin{align}
\frac{\ud}{\ud t}\Bigl\langle\prod_{i\in\mathcal{A}}\sigma_i^{\mu_i}\Bigr\rangle \;\;\;=&\;\;\; \sum_{i\in\mathcal{A}}\varepsilon^{\mu_i\alpha\nu}h_i^\alpha\Bigl\langle\sigma_i^\nu\prod_{j\in\mathcal{A}\backslash\{i\}}\sigma_j^{\mu_j}\Bigr\rangle\;\;+\;\; \sum_{i\in\mathcal{A}}\sum_{\ell\not\in\mathcal{A}}\varepsilon^{\mu_i\alpha\nu}V_{i\ell}^{\alpha\lambda}\Bigl\langle\sigma_\ell^\lambda\sigma_i^\nu\prod_{j\in\mathcal{A}\backslash\{i\}}\sigma_j^{\mu_j}\Bigr\rangle\nonumber\\
& \qquad\qquad\qquad\qquad + \;\; \sum_{i\in\mathcal{A}}\sum_{j\in\mathcal{A}\backslash \{i\}}\varepsilon^{\mu_i\alpha\nu}V_{ij}^{\alpha\mu_j}\Bigl\langle\sigma_i^\nu\prod_{k\in\mathcal{A}\backslash\{i,j\}}\sigma_k^{\mu_k}\Bigr\rangle.\label{eq:apeomc}
\end{align}
\end{widetext}
which is the hierarchy of equations of motion for the spin correlators that we wished to derive (cf. eqtn. (\ref{eq:EOMgensp12})).


\section{Matrix Propagator for 2-spin system}
 \label{sec:App-C}


In the main text we worked out explicitly the equation of motion for the entanglement correlators of a simple 2-spin system, with the Hamiltonian 
\begin{equation}
H=\tfrac{1}{2} \left[ \Delta_1 \tau^x+\Delta_2\sigma^x+\omega\tau^z\sigma^z \right]
 \label{eq:HDeltaomega2'}
\end{equation}
and eigenvalues $\epsilon_1$, $\epsilon_2$ (compare eqtn. (\ref{eq:HDeltaomega2}) {\it et seq.}). 

Here we write out explicitly the propagators which appear in the block matrix $\mathbb{G}(z)$ (the result for $\mathbb{G}(t)$ then being given by Fourier transformation). We have
\begin{widetext}
\begin{align}
\vc{g}_{11}(z)\;=&\;\left(\frac{\omega^2z}{2\omega_{30}(z^2+\omega_{30}^2)}+\frac{\omega^2z}{2\omega_{21}(z^2+\omega_{21}^2)}+\left(1-\frac{\omega^4}{\omega_{30}^2\omega_{21}^2}\right)\frac{1}{z}\right)\hat{x}\hat{x}\nonumber\\
&+\frac{z}{2\omega_{30}\omega_{21}}\left\{\left(\frac{\omega_{20}^2-\Delta_1^2}{z^2+\omega_{10}^2}+\frac{\omega_{20}^2-\Delta_2^2}{z^2+\omega_{20}^2}\right)(\hat{y}\hat{y}+\hat{z}\hat{z})-2\omega^2\left(\frac{\hat{y}\hat{y}}{z^2+\omega_{10}^2}+\frac{\hat{z}\hat{z}}{z^2+\omega_{20}^2}\right)\right\}\nonumber\\
&+\frac{\hat{y}\hat{z}-\hat{z}\hat{y}}{4\omega_{30}\omega_{20}}\left\{\left[\omega_{21}(\Delta_1+\Delta_2)-\omega_{30}(\Delta_1-\Delta_2)\right]\frac{\omega_{10}^2}{z^2+\omega_{10}^2}+\left[\omega_{21}(\Delta_1-\Delta_2)+\omega_{30}(\Delta_1+\Delta_2)\right]\frac{\omega_{20}^2}{z^2+\omega_{20}^2}\right\}\\
\vc{g}_{12}(z)\;=&\;\;\hat{x}\hat{x}\left\{\frac{2\Delta_1\Delta_2\omega^2}{\omega_{30}^2\omega_{21}^2z}+\frac{z\omega^2}{2}\left(\frac{1}{\omega_{30}(z^2+\omega_{30}^2)}-\frac{1}{\omega_{21}(z^2+\omega_{21}^2)}\right)\right\}\;\;=\;\; \vc{g}_{21}(z)\\
\end{align}
for the ``small" matrix propagators, and 
\begin{align}
{g}_{1p}^{\mu}{}_{\nu\beta}=&\frac{\Delta_1\omega\left(2\Delta_2z\hat{x}^\mu\hat{z}_\nu\hat{y}_\beta+[z^2+\Delta_1^2-\Delta_2^2+\omega^2]\hat{x}^\mu\hat{z}_\nu\hat{y}_\beta-[z^2+\Delta_1^2+\Delta_2^2+\omega^2]\hat{x}^\mu\hat{y}_\nu\hat{z}_\beta-\tfrac{\Delta_2}{z}[z^2-\Delta_1^2+\Delta_2^2+\omega^2]\hat{x}^\mu\hat{y}_\nu\hat{y}_\beta\right)}{z(z^2+\omega_{21}^2)(z^2+\omega_{30}^2)}\nonumber\\&\quad\quad+\omega\frac{z\Delta_2\hat{y}^\mu\hat{x}_\nu\hat{y}_\beta+z^2\hat{y}^\mu\hat{x}_\nu\hat{z}_\beta+\Delta_1\Delta_2\hat{z}^\mu\hat{x}_\nu\hat{y}_\beta+z\Delta_1\hat{z}^\mu\hat{x}_\nu\hat{z}_\beta}{z^2\omega^2+(z^2+\Delta_1^2)(z^2+\Delta_2^2)}\\
{g}_{pp}^{\mu\alpha}{}_{\nu\beta}=&\frac{1}{z}\hat{x}^\mu\hat{x}^\alpha\hat{x}_\nu\hat{x}_\beta  +  \left[(z^2+\Delta_1^2)(z^2+\Delta_2^2)+z^2\omega^2\right]^{-1}\biggl\{z(z^2+\Delta_1^2+\omega^2)\hat{x}^\mu\hat{y}^\alpha\hat{x}_\nu\hat{y}_\beta+\Delta_2(z^2+\Delta_1^2)\left[\hat{x}^\mu\hat{z}^\alpha\hat{x}_\nu\hat{y}_\beta-\hat{x}^\mu\hat{y}^\alpha\hat{x}_\nu\hat{z}_\beta\right]\nonumber\\
&+z(z^2+\Delta_1^2)\hat{x}^\mu\hat{z}^\alpha\hat{x}_\nu\hat{z}_\beta+z(z^2+\Delta_2^2+\omega^2)\hat{y}^\mu\hat{x}^\alpha\hat{y}_\nu\hat{x}_\beta+\Delta_1(z^2+\Delta_2^2)\left[\hat{z}^\mu\hat{x}^\alpha\hat{y}_\nu\hat{x}_\beta-\hat{y}^\mu\hat{x}^\alpha\hat{z}_\nu\hat{x}_\beta\right]\nonumber\\
&+z(z^2+\Delta_2^2)\hat{z}^\mu\hat{x}^\alpha\hat{z}_\nu\hat{x}_\beta\biggr\}+(z^2+\omega_{21}^2)^{-1}(z^2+\omega_{30}^2)^{-1}\biggl\{(z^2+\Delta_1^2+\Delta_2^2+\omega^2)\Bigl[\left(z+\tfrac{\omega^2}{z}\right)\left(\hat{y}^\mu\hat{y}^\alpha\hat{y}_\nu\hat{y}_\beta+\hat{z}^\mu\hat{z}^\alpha\hat{z}_\nu\hat{z}_\beta\right)\nonumber\\&+z\left(\hat{y}^\mu\hat{z}^\alpha\hat{y}_\nu\hat{z}_\beta+\hat{z}^\mu\hat{y}^\alpha\hat{z}_\nu\hat{y}_\beta\right)\Bigr]+\Delta_2(z^2-\Delta_1^2+\Delta_2^2+\omega^2)\left[\hat{y}^\mu\hat{z}^\alpha\hat{y}_\nu\hat{y}_\beta-\hat{y}^\mu\hat{y}^\alpha\hat{y}_\nu\hat{z}_\beta+\hat{z}^\mu\hat{z}^\alpha\hat{z}_\nu\hat{y}_\beta-\hat{z}^\mu\hat{y}^\alpha\hat{z}_\nu\hat{z}_\beta\right]\nonumber\\&
+\Delta_1(z^2+\Delta_1^2-\Delta_2^2+\omega^2)\left[\hat{z}^\mu\hat{y}^\alpha\hat{y}_\nu\hat{y}_\beta+\hat{z}^\mu\hat{z}^\alpha\hat{y}_\nu\hat{z}_\beta-\hat{y}^\mu\hat{y}^\alpha\hat{z}_\nu\hat{y}_\beta-\hat{y}^\mu\hat{z}^\alpha\hat{z}_\nu\hat{z}_\beta\right]\nonumber\\
&+2\tfrac{\Delta_1\Delta_2}{z}(z^2+\omega^2)\left[\hat{y}^\mu\hat{y}^\alpha\hat{z}_\nu\hat{z}_\beta-\hat{z}^\mu\hat{z}^\alpha\hat{y}_\nu\hat{y}_\beta\right]+2z\Delta_1\Delta_2\left[\hat{y}^\mu\hat{z}^\alpha\hat{z}_\nu\hat{y}_\beta-\hat{z}^\mu\hat{y}^\alpha\hat{y}_\nu\hat{z}_\beta\right]\biggr\}.
\end{align}
\end{widetext}
for the ``large" matrix propagators. In these equations $\hat{x}, \hat{y}$, and $\hat{z}$ are unit Cartesian vectors, and $\hat{z}$ should not be confused with the complex frequency $z$.

Formulae for $\vc{g}_{22},\, \vc{g}_{21},$ and $\vc{g}_{2p}$, can be obtained from the expressions  for $\vc{g}_{11},\, \vc{g}_{21},$ and $\vc{g}_{1p}$, if we make the replacements $\Delta_1\to\Delta_2$ and $\Delta_2\to\Delta_1$ and adjust the tensor indices accordingly ($\mu\to\alpha,\,\nu\to\beta,\,\alpha\to\mu,\,\beta\to\nu$). $\vc{g}_{p1}(z)$ and $\vc{g}_{p2}(z)$ can be obtained from $\vc{g}_{1p},\vc{g}_{2p}$ using the identities $\vc{g}_{p1}(z)=\vc{g}_{1p}^T(-z)$ and $\vc{g}_{p2}(z)=\vc{g}_{2p}^T(-z)$ (we have obtained these identities by examining the full solution).


\end{document}